\RequirePackage{fix-cm}
\documentclass[smallextended]{svjour3}       
\smartqed  
\usepackage{graphicx}
\usepackage{subfigure}
\usepackage{url}
\usepackage{amsmath}

\usepackage[usenames, dvipsnames]{color}

\begin{document}

\title{$C^3$-index: A PageRank based multi-faceted metric for authors' performance measurement 
}

\titlerunning{$C^3$-index for authors' performance measurement}        

\author{Dinesh Pradhan       \and
        Partha Sarathi Paul   \and
        Umesh Maheswari		  \and
        Subrata Nandi		  \and
        Tanmoy Chakraborty$^*$
}


\institute{Dinesh Pradhan, Partha Sarathi Paul, Umesh Maheswari, Subrata Nandi \at
              Dept. of Computer Science \& Engineering, National Institute of Technology, Durgapur, India \\
              \email{dineshkrp@gmail.com,mtc0113@gmail.com,umeshmaheswari7@gmail.com,subrata.nandi@gmail.com}          
           \and
           Tanmoy Chakraborty \at
              Dept. of Computer Science, University of Maryland, College Park, USA \\
              \email{tanchak@umiacs.umd.edu}\\
              $^*$ Corresponding author
}

\date{Received: date / Accepted: date}

\maketitle

\begin{abstract}
Ranking scientific authors is an important but challenging task, mostly due to the dynamic nature of the evolving scientific publications. The basic indicators of an author's productivity and impact are still the number of publications and the citation count (leading to the popular metrics such as h-index, g-index etc.). H-index and its popular variants are mostly effective in ranking highly-cited authors, thus fail to resolve ties while ranking medium-cited and low-cited authors who are majority in number. \textcolor{black}{Therefore, these metrics are inefficient to predict the ability of promising young researchers at the beginning of their career.} In this paper, we propose $C^3$-index that combines the effect of  citations and collaborations of an author in a systematic way using a weighted multi-layered network to rank authors. We conduct our experiments on a massive publication dataset of Computer Science and show that -- (i) $C^3$-index is consistent over time, which is one of the fundamental characteristics of a ranking metric, (ii) $C^3$-index is as efficient as h-index and its variants to rank highly-cited authors, (iii) $C^3$-index can act as a conflict resolution metric to break ties in the ranking of medium-cited and low-cited authors, \textcolor{black}{(iv) $C^3$-index can also be used to predict future achievers at the early stage of their career.}

\end{abstract}


\section{Introduction}
\label{intro}
\begin{quotation}
``...which indices are preferred depends on the question that is asked. No single index provides an optimal metric of science, whether scaled at the level of the individual scientist, topic, field, journal, or discipline.''
\end{quotation}
\hfill John T. Cacioppo \cite{metric}

\textcolor{black}{How do we quantify the quality of science? The question is neither rhetorical nor an emotional one; it is very much relevant to promotion committees, funding agencies, national academies, politicians and so on, in order to recognize and acknowledge quality research and prominent researchers. Identifying high-quality research is necessary for the advancement of science, but measuring the quality of research is even more important in today's world when scientists in different research fields are increasingly competing with each other for different purposes such as receiving research grants, publishing papers in prestigious vanues (conferences/journals) etc. The widely accepted approach is to check the bibliographic record of a researcher -- that is, the number and the impact of publications. Researchers with very different bibliographic credentials may have the same h-index \cite{bornmann2009state}. Kosmulski pointed out that h-index is more suitable for the assessment of mature scientists who have published at least 50 papers and have h-indexes of at least 10 \cite{Kosmulski_2006_ISSI}.}


\textcolor{black}{Assessment of science is important for many different reasons. For researchers at an early stage of their careers, a metric of scientific work may provide significant feedback to their progress and their exact position in the scientific world. For the recruitment committees in universities/research institutes, such a metric may simplify the task of wading through bunch of applications to select a list of potential applicants for the interview. For university administrators, these metrics may help to judge researchers seeking promotion or tenure. For the departmental chairs in an Institute, these metrics may help suggesting annual raises and the allocation of scarce departmental resources. For scientific societies, these metrics may influence selecting award recipients. For research granting agencies, an assessment of scientific fields would help identifying areas of progress and vitality. For legislative bodies and boards of directors, a measure of science may provide a means of documenting performance, ensuring accountability, and evaluating the return on their research investment. Measures of science may have other applications such as identifying the structure of science, the impact of academic journals, influential fields of research in current time, and factors that may contribute to new discoveries \cite{cacioppo2016social}.}


Several studies have been conducted by formulating the scientific progress in terms of networks (such as citation network, coauthorship/collaboration network) \cite{Pradhan0PN16,Pradhanwebsci,0002SGM14,0002GM15,0002GM14}. Studies on coauthorship networks focus on network topology and network statistical mechanics \cite{centrality_measures_to_impact_analysis_JASIST}. Although our research also deals with citation and collaboration networks, we take a different approach by studying micro-level network properties, with the aim of applying centrality measures such as PageRank for impact analysis \cite{centrality_measures_to_impact_analysis_JASIST}.

In citation analysis, the number of citations reflects the impact of a scientific publication. This measurement considers each citation equally --- a citation coming from an obscure paper has the same weight as one from a ground-breaking, highly-cited work \cite{maslov2008promise}. Pinski and Narin \cite{pinski1976citation} were the first to note the difference between popularity and prestige in the bibliometric area. They proposed using the eigenvector of a journal citation matrix (i.e., similar to PageRank) corresponding to the principal eigenvalue to represent journal prestige. \textcolor{black}{Bollen et al. \cite{Bollen2006} defined journal prestige and popularity, and developed a weighted PageRank algorithm to measure them.} They defined `popular' journals as those which are cited frequently by journals with little prestige, and `prestigious' journals as those which are cited by highly prestigious journals. Their definitions are recursive. Recently, Ding and Cronin \cite{popular_or_prestigious} extended this approach to authors and applied weighted citation count to measure \textcolor{black}{researcher's} prestige in the field of information retrieval. They defined the popularity of a researcher as the number of times he/she is cited (endorsed) in total, and prestige as the number of times he/she is cited by highly cited papers. \textcolor{black}{The main idea behind this prestige measure is to use simple citation count but to give more weights to highly-cited papers.}

Since scholarly activities are often represented in the form of complex networks where authors, journals, and papers are connected via citing/being cited or coauthored, the network topology can significantly influence the impact of an author, journal, or paper. The recent developments of large-scale networks and the success of PageRank demonstrate the influence of the network topology on scholarly data analysis. \textcolor{black}{PageRank or weighted PageRank have performed well in representing the prestige of journals \cite{Bollen2006,falagas2008comparison}; however relatively few researchers have applied this concept to authors. Some of the works that addressed the issues include Fiala et al. \cite{Fiala2008}, Ding \cite{pagerank_to_author_citation_JASIST}, Radicchi et al. \cite{Diffusion_of_Scientific_Credits}, {\.Z}yczkowski  \cite{zyczkowski2010citation} etc. These papers are built following the notion of Ding and Cronin \cite{popular_or_prestigious} and clearly address the issue why PageRank or weighted PageRank based algorithms could be applied to author citation networks for measuring the popularity and the prestige of scholars.}

In this paper, we use citation networks of authors, publications and journals, constructed from  a massive publication dataset related to Computer Science domain. Our aim is to find a measure with which one can rank the authors of scientific papers appropriately. Our proposed method includes the adoption of the PageRank algorithm, which can be considered as a measure of prestige, as well as a measure of significance. 


\textcolor{black}{Recently, through a bibliometric analysis of the entire Italian university population working in the hard sciences over the period 2001 - 2005, Abramo et al. \cite{Are_Performers_collaborators} attempted to answer some of the questions related to bibliographic research.} The results show that the researchers with top performance with respect to their national colleagues are also those who collaborate more abroad; but that the reverse is not always true. Collaboration is a fundamental aspect of scientific research activity. The reasons for collaboration are many, however most can probably be attributed to a ``pragmatic attitude to collaboration" \cite{melin2000pragmatism}.

In our present work, we propose an author performance metric called $C^3$-index that ranks authors based on their received citations as well as their collaboration profile through a PageRank based strategy. The proposed index has moderate correlation (60\%) with h-index. It is observed that one of the component scores (ACI-score) of the proposed index has very strong correlation (98\%) with h-index, but the other two component scores (PCI-score and AAI-score) have significantly less correlation (40\% - 50\%). These observations suggest that the proposed index carries more information than h-index. We further observe that a significant fraction of authors having high AAI-score during the start of the time-frame (1998 - 2008) have achieved significantly high h-index  during the end of the time-frame. We also notice that the authors who we find reaching a certain performance level in terms of their h-index values during the end of a given time-frame reach moderately high performance level according to the proposed index at least 4-5 years in advance. This observation indicates the future prediction capability of the proposed index as well.

\begin{figure}
  \centering
  \subfigure[]{ 
  \includegraphics[scale=.21]{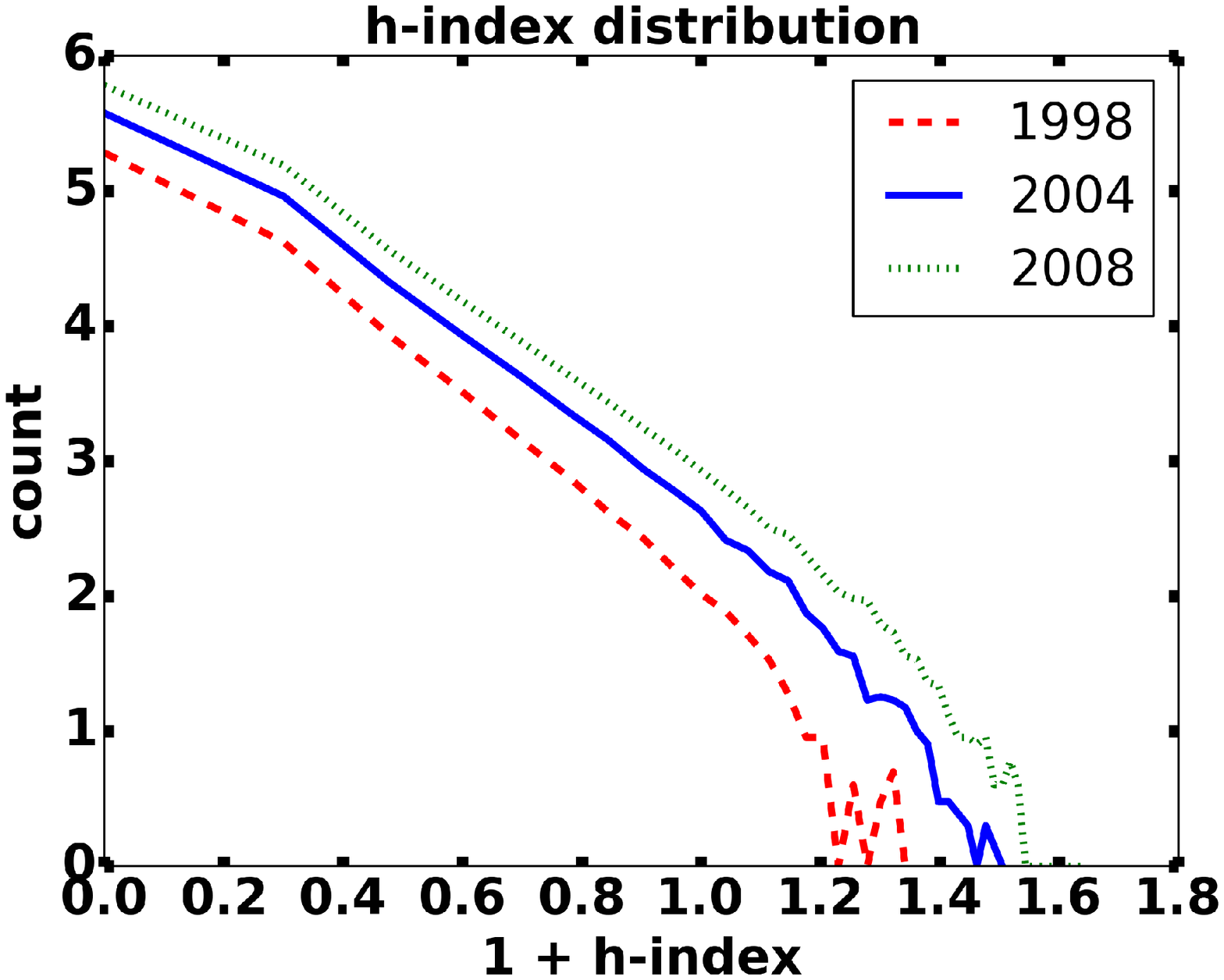}
  \label{figure:h_distribution_log_log}
   }
  \subfigure[]{
  \includegraphics[scale=.21]{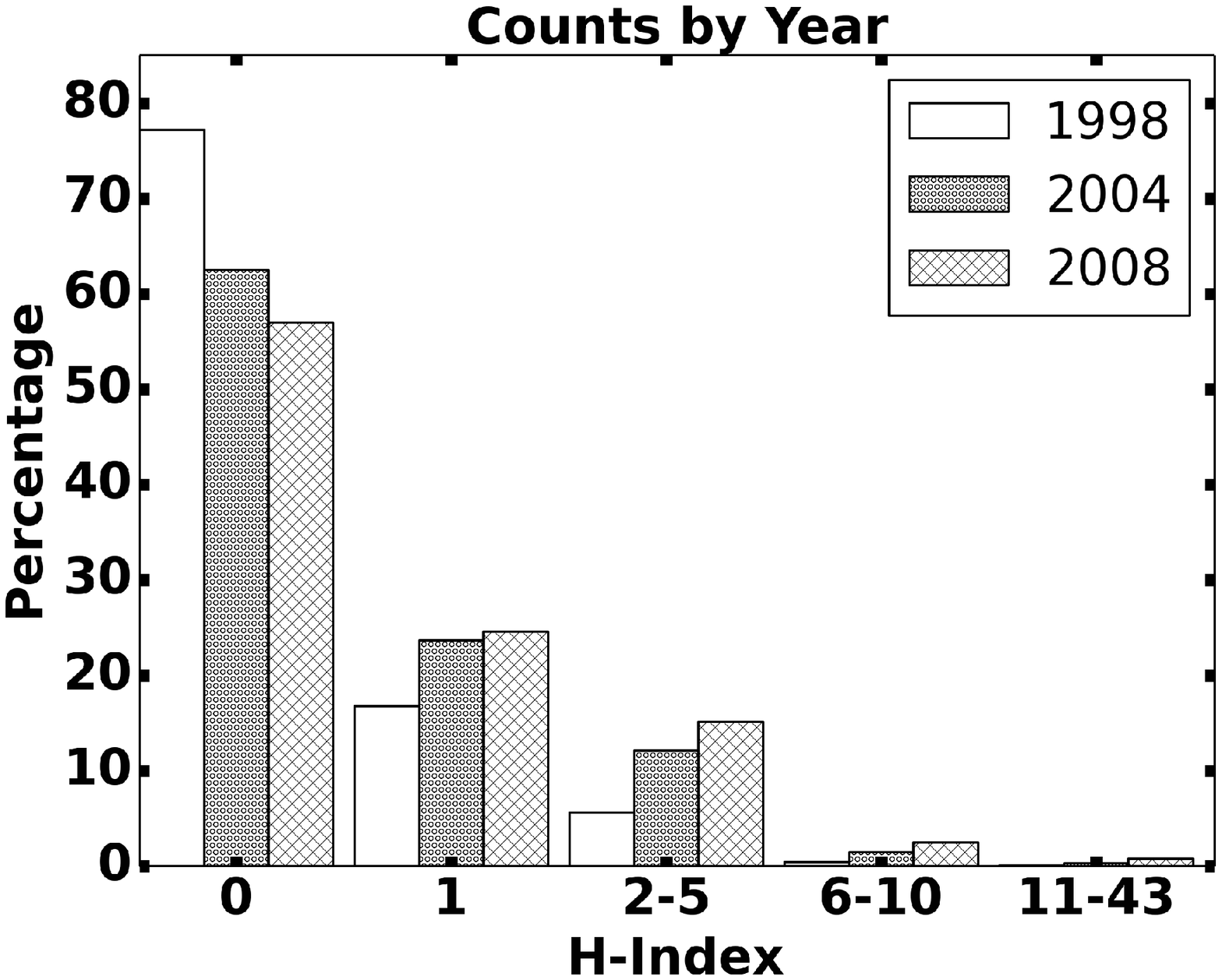}
  \label{figure:h_distribution_barplot}
   }
  \subfigure[]{ 
  \includegraphics[scale=.21]{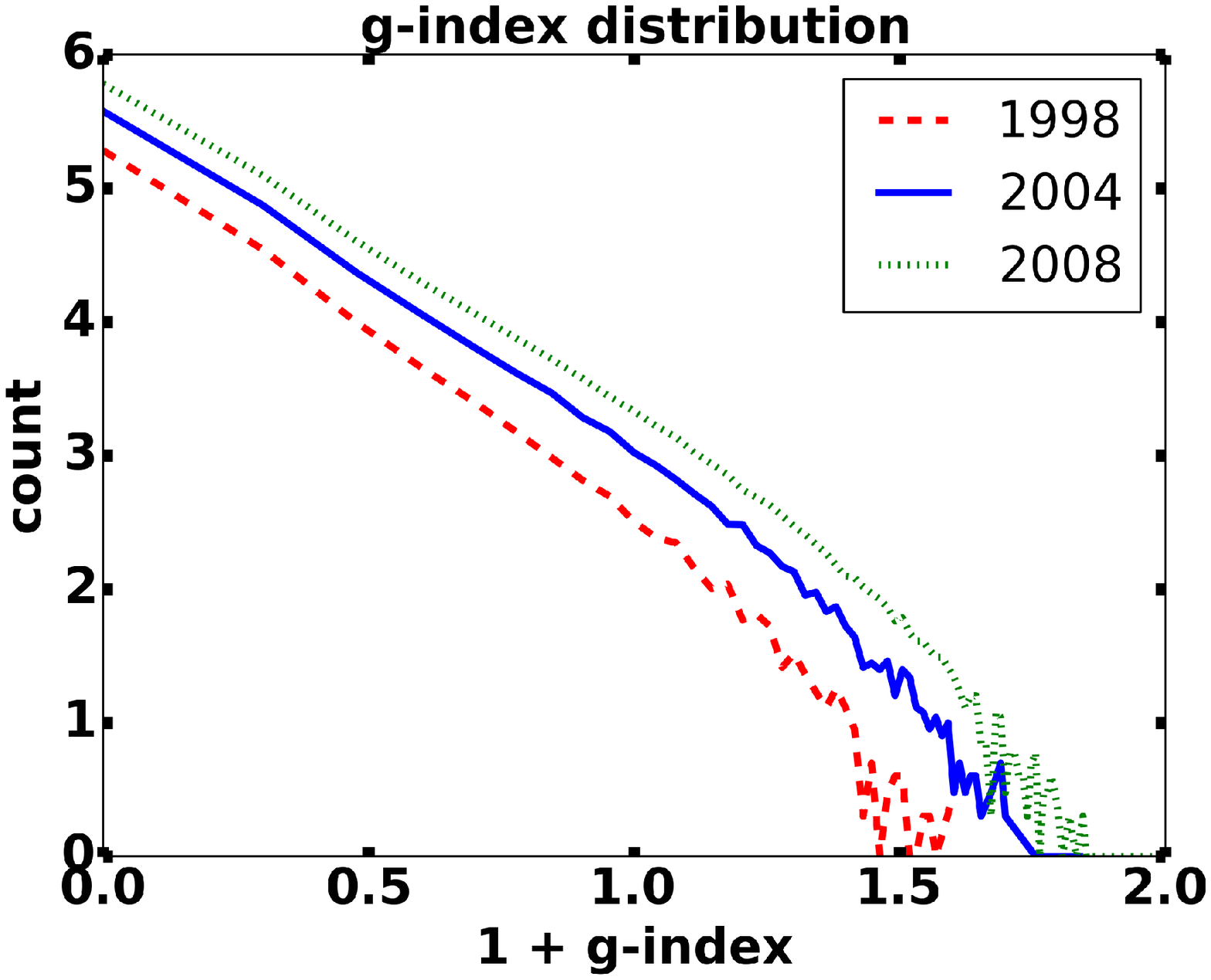}
  \label{figure:g_distribution_log_log}
   }
  \subfigure[]{ 
  \includegraphics[scale=.21]{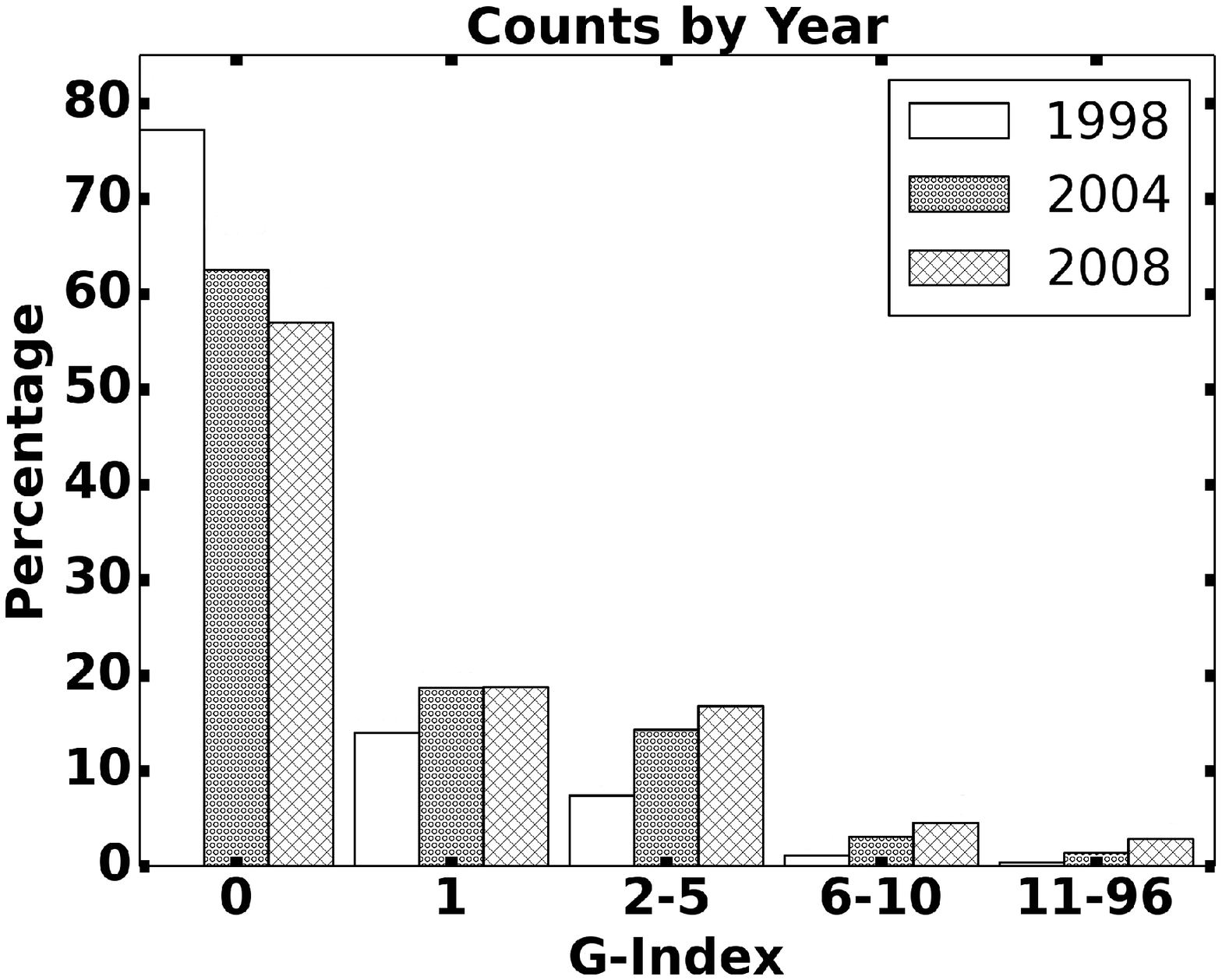}
  \label{figure:g_distribution_barplot}
   }
  \caption{(a) Number of authors are plotted against different h-index values (plus one) in log-log scale for three different years: 1998, 2004 and 2008. (b) Percentage of authors as distributed across five different h-index bins for the year 1998 (left bars), distribution of the same set of authors in 2004  (middle bars), and in 2008 (right bars). The figure reflects that a limited fragment of authors attain high h-index over the years, but majority remains unimproved. (c) and (d) show the similar plots against g-index. The near straight line nature of all the curves in (a) and (c) ensures power-law behavior of both h- and g-index. \textcolor{black}{(b) and (d) suggest that a small fragment of authors having low index values gradually improve over the years, whereas the majority remain unchanged. It is necessary to characterize as well as to predict, in well advance, the fragment of authors that have prospect of improvement.}}
\label{figure:First_Plot}
\end{figure}

\section{\textcolor{black}{Related Work}}
\label{related-work}

\textcolor{black}{To propose strategies for ranking authors, researchers from citation analysis and other domains largely use publications made by the corresponding authors and the citations received by those publications.} The seminal work by \emph{J. E. Hirsch} proposed \emph{h-index}~\cite{h-index} considering both the number of publications and citations in a balanced way. H-index and its variants gained wide acceptance in the research community {\color{black} because they are easy to compute}, \textcolor{black}{though critics pointed out their limitations~\cite{Advantages_Limitations_of_h-index,Inconsistencies_of_h-index,Limitations_H-index}}.

\textcolor{black}{Analysis of publication trajectory of authors from the faculty of psychology during their first seven years of post-doctoral studies revealed that the rate of publication increased each year following completion of their doctorate program \cite{byrnes2007publishing}. The largest rate of increase in publication counts of peer-reviewed journal articles was observed in the first four years rather than in the two years immediately before their tenure. Publication count prior to tenure does not tell the whole story, of course. In an investigation of gender differences in scientific productivity, Long \cite{long1992measures} found that women publish fewer articles than men during the first decade of their career, but this difference is reversed later in their careers. According to a search of the ISI database, John Ridley Stroop published only three papers during his career. The articles has been cited 3,810 times, whereas other two papers received less than 1 percent of the citations of former paper \cite{metric}. Total number of citations is necessary for the evaluation of one's scientific merit, but it misses the point that Stroop's scientific contributions to psychology were limited primarily to his efforts prior to the completion of his PhD. Any good metric for scientific qualities of a researcher should capture in its quantification such instantaneous rise and fall of an author during her research career, which in our opinion, no naive metric for scientific impact is capable of.}

Many alternative proposals were made in the line of h-index to overcome those limitations, as well as to use the power of h-index --  Hirsch himself proposed $\hbar$-index (pronounced as \emph{ $hbar$-index})~\cite{hbar-index} that considers multiple coauthors of a paper that h-index omits; Egghe et al. proposed \emph{g-index}~\cite{g-index}. \textcolor{black}{Jin et al. \cite{Jin2007} proposed {\em AR-index}, another interesting metric whose value may decrease over time due to aging of citations. This somehow could downgrade researchers who `rest on their laurels' for long.}


\textcolor{black}{H-index and its variants are elegant as well as have a concrete mathematical foundation \cite{Meaning_of_h_index} -- these are integral measures and have narrow bounds (order of hundreds)\footnote{http://www.webometrics.info/en/node/58}}. On the other hand, the number of authors engaged in active research nowadays are in the range of millions. This infers, through \textit{pigeon-hole principle} that every single h-index instance is associated with a very large number of authors. Again, there are very few authors having high h-index, and most of the authors lie in low h-index region. So, we may expect a power-law  behavior in the distribution of h-index (and its variants such as g-index) across author spectrum, which is shown in Figures \ref{figure:h_distribution_log_log} and \ref{figure:g_distribution_log_log}). Narrow resolution for the middle and bottom liners in case of citation-count based indices restricts the research community from any fine-grained analysis of the authors residing in that part of the spectrum, which is very much essential for predicting the future position; even a Nobel laureate has to start his/her career with very low h-index. \textcolor{black}{As observed from Figures \ref{figure:h_distribution_barplot} and \ref{figure:g_distribution_barplot},  a significant mass of the bottom liners gradually attain high h-index or g-index, whereas the remaining mass is nearly static over time.}

H-index and its variants use only the citations received by individual papers of the concerned author for ranking the authors, though there are other features of an author that influence his/her career. Abramo et al. tested through an investigation on Italian university system that the authors who collaborate more at international level perform better than those who collaborate less; though the converse, they also observed, is not true in general~\cite{Are_Performers_collaborators}. Credit sharing among coauthors of a multi-authored scientific paper is still an  unresolved issue, though some attempts were made in this context. Trueba et al. proposed a robust formula for the same \cite{Formula_to_Credit_Authors}, where the credit is shared among coauthors based on their relative position in the author name sequence in a paper. \textcolor{black}{However such a distribution may not be fool-proof, and may not be applicable to some domains\footnote{https://en.wikipedia.org/wiki/Academic\_authorship}, where maintaining strict alphabetical name sequence is a common practice. Other concise study on the topic was undertaken by Xu et al. \cite{Survey_Author_Credit_Assignment_Schemas} and Tscharntke et al. \cite{Credit_for_Contribution_PLOS}, where they categorically discussed and analysed different proposed schemes, and pointed out the lack of any conclusive decision.} A consensus about the credit sharing among coauthors is expected, because a promising star usually starts his/her career as a primary coauthor of an existing star; however the converse may not be true. Other influential factors may be the affiliation they have, the venues they choose for publication, the countries they belong to, and so on~\cite{Factors_Affecting_Citations}. 

As an alternative, some PageRank based schemes for ranking papers and authors have also been proposed. \textcolor{black}{Chen et al.~\cite{finding_scientific_gems_using_google_pagerank} used Google PageRank method on the citation network formed by articles published in the Physical Review journals with the goal of measuring the importance of an individual scientific publication. They also pointed out that a choice of $0.5$ as the damping factor suits better in the present context as compared to $0.85$ in case of conventional webpage hyperlink network.} Ma et al.~\cite{pagerank_to_Citation_Analysis} claimed that a PageRank based representation might be a better indicator serving as a substitution of the number of citations for measuring the influence of a paper. \textcolor{black}{Ding et al. \cite{popular_or_prestigious} differentiated the scholarly popularity versus the scholarly prestige of an author -- the popularity of an author is the number of times the author is referenced by other papers, whereas the prestige is the number of times the author is cited by only highly-cited papers.}

\textcolor{black}{PageRank based methods for ranking authors use one or more from variety of features -- citations from the peer colleagues, coauthorship with other researchers, co-citations for the papers/authors, and so on.} Radicchi et al.~\cite{Diffusion_of_Scientific_Credits} proposed an author ranking algorithm based on diffusion of scientific credits using a proposed weighted author citation network. Barab{\'a}si et al. \cite{Social_Network_of_Scientific_Collaboration_Barabasi} studied the coauthorship among researchers as a social network and observed its dynamic behavior over time. Ortega \cite{influence_of_coauthorship_on_research_impact_MAS} observed that the structure of an author co-authorship network may reveal a good deal of information about research performance of a researcher through the analysis of the data from Microsoft Academic Search.  Liu et al.~\cite{coauthorship_netwrok_in_DL_research_community} studied the co-authorship network of the Digital Library (DL) research community as represented in the ADL, DL and JCDL conference series to reveal the structure of collaborations within the DL research community and the quantitative metrics for the concepts of status and influence. They built a weighted and directed network model to represent collaboration relationships, and proposed ``AuthorRank'', an alternative metric for ranking authors' prestige. Ding et al. \cite{pagerank_for_ranking_authors} observed the effects of different damping factors (ranging from 0.05 to 0.95) for author co-citation network, and noted that citation rank is close to PageRank for damping factor of 0.55. \textcolor{black}{A detailed study on PageRank variants for ranking authors may be found in the  work by Michal Nykl et al.~\cite{pagerank_variants_in_citation_networks}.}

\emph{What would happen if one prefers to use more than one factor at a time in a PageRank based approach?} \textcolor{black}{In a very recent work \cite{Senanayake_PLOS_2015} and two of its preceding works \cite{Senanayake_IEEE_2014,Senanayake_Elsevier_2014}, Senanayeke et al. proposed {\em PageRank-Index} ({\em aka} {\em p-index} sometimes) that ranks authors by a PageRank-based approach using paper-paper citation and author-author coauthorship features at the same time. In their approach, the score for each paper is calculated using a PageRank-based approach; the score for each paper is distributed among all the coauthors of the paper in a weighted manner, where the weights are determined by the order of their authorship in the corresponding paper; finally, the PageRank shares obtained as above is summed and a percentile score is computed from the sum to get the final PageRank score. The approach is very effective, except a couple of points as follows -- (a) the order of authorship may not be a standard measure of author contribution in a paper, as mentioned earlier, (b) the proposed approach does not take advantage of the network property of author-author coauthorship network, and thus may miss some of the greater insights that the network could provide. In our work, we use network properties of author-author citation network and author-author coauthorship network for redistributing the PageRank based paper scores among the coauthors of the papers. To achieve this, we apply PageRank-based computation on these two layers (paper-paper citation network and author-author coauthorship network) as well.}

In a student project at Stanford University, Cui et al. \cite{citation_network_as_multilayer_graph} proposed to represent the citation network as a multilayer network, which is the first attempt towards modeling multiple factors together for scholarly research impact metric. The technical report by S. Boccaletti et al.~\cite{structure_multilayer_networks} explains different aspects and applications of multilayer networks. It explicitly shows the areas where monoplex networks fail to capture the full detail of the scenario; whereas the multilayer networks may provide a better insight. Halu et al.~\cite{multiplex_pagerank} proposed an idea of biased random walks to define the PageRank centrality measure on multiplex networks. Domenico et al.~\cite{nature_ranking_in_multilayer} claimed that calculating the centrality of nodes in component networks of the multilayer structure separately or aggregating the information to a single network leads to misleading results. They proposed to use tensorial formulation of multilayer networks to overcome the limitations.

\textcolor{black}{There is a series of research that uses {\em heterogeneous networks}, which, in our observation, are very similar to multilayer networks.} Zhou et al.~\cite{Zhou_Author_Ranking} used a heterogeneous network (similar to a two-layer network) for co-ranking authors and documents simultaneously. The co-ranking framework they adopted uses intra- and inter-class random walks to design a PageRank based strategy on heterogeneous networks. Expert finding with particular type of expertise for a given query is a non-trivial task, especially from a large-scale web systems, such as question answering and bibliography data, and is very much similar to the objective like author ranking. Deng et al.~\cite{Heterogeneous_expertize_ranking} proposed a joint regularization framework to enhance expert retrieval by modeling heterogeneous networks as regularization constraints on top of document-centric model. Yan et al.~\cite{p-rank} used heterogeneous network similar to three-layer network model for ranking authors.

\emph{Do PageRank based strategies reveal more information than simple citation-count based approaches?} \textcolor{black}{Recently, Fiala et al.~\cite{Fiala2015334} claimed that there is no evidence that PageRank based approaches certainly outperform simple citation-count based ranking approaches.} The motivation of our  work stems from this conclusion -- we would like to design a PageRank-based ranking scheme that can provide additional information which may not be obtained from simple citation-count based strategies.

\section{Motivation}
\label{motivation}

From the existing literature on author ranking strategies, one may observe that the following features are used for ranking authors: (a) paper-paper citation~\cite{h-index,g-index,hbar-index}, (b) author-author citation~\cite{popular_or_prestigious}, (c) author-author cocitation~\cite{pagerank_for_ranking_authors}, (d) author-author coauthorship~\cite{influence_of_coauthorship_on_research_impact_MAS,coauthorship_netwrok_in_DL_research_community}, \textcolor{black}{(e) author-author collaboration\footnote{A supergraph of author-author coauthorship graph that takes into account social relationship between authors other than coauthorship: friends in the social media, Committee members of the same conference, editors of the same journal, members having same affiliation, etc. However, this feature is not frequently used due to the lack of suitable dataset.} \cite{Murthy01012015}}.

Citations from other papers are possibly a well-accepted measure of the influence of a paper in the bibliographic domain. On the other hand, an author is best judged by the publications he/she made during his/his research life. A large pool of existing research follows this simple reasoning, and uses only \textit{paper-paper citation network} to rank authors. \textcolor{black}{However, not all papers of an author are of the same stature, and hence have different ranks.} Now, combining these individual paper scores to a consistent author score might be challenging as well as debatable. An alternative solution might be to derive author rank solely from the author network; \textcolor{black}{and possibly the first point to assume in this category is the one derived from paper-paper citations, viz, \textit{author-author citation network}. Another, slightly less-intuitive is author-author cocitation networks, where two authors are connected if they cite the same set of papers.} An intuitive justification might be that authors working on the same topic usually read and refer to the same set of papers.  

One fundamental limitation associated with citation-based scoring technique is that it takes some time to gain attention after it is eventually published~\cite{Tanmoy_Citation_Dynamics}. Also the time required for a paper to be published after it is actually communicated to a venue is not small. Due to this factor, recent publications usually are misjudged if they are indexed only on the basis of citation count. \textcolor{black}{The same limitation is observed in case of young authors, for whom major publications are quite recent and may not get enough attention in the early days.}

To get rid of these limitations, other features such as author-author coauthorship or author-author collaboration were tried \cite{centrality_measures_to_impact_analysis_JASIST}. The reasons might be that high performers usually collaborate with other high-performers, either in case of coauthorship or in case of social collaboration. As an example, entry of a research student under an eminent researcher's supervision is quite a hurdle; so are the Technical Program Committee (TPC) members in a good conference, or to be an editor in a prestigious journal. \textcolor{black}{We can thus assume that coauthors of an eminent researcher  more or less have same calibre; and the same could be assumed for the TPC members of a good conference, or editors of a reputed journal.} Hence, it is quite reasonable to exploit such social relationships to derive the influence of an author in the respective author network.

\textcolor{black}{However, citation profile is a prominent feature to measure the influence of a paper, and hence removing it completely from the scope of study is unjustified.} A better approach might be to devise a scoring strategy that considers all the above features mentioned earlier. \textcolor{black}{However, we see that each individual feature leads to a complex network, directed or undirected, combining which leads to a multilayer complex network.} It is quite evident that a PageRank like computation on a multilayer complex network is cost-inefficient. So the reduction of possible redundancy in the feature set, leading to the reduction of dimension of the underlying complex network, might save a huge amount of computation. If the resulting feature set (after removal of possible feature level redundancy) consists of only one feature, the resulting complex network would be a single layer network. \textcolor{black}{Otherwise, we would try to find a minimal feature set that would lead to a multilayer complex network with least dimensions.}

Through the reasoning so far, we find that author-author citation and author-author coauthorship are two indispensable features for any study for ranking authors. However, though an author-author citation network can be derived from a  paper-paper citation  network, the latter neither could replace the former, nor can be removed completely. The reason is that an author may not be judged completely without considering the quality of individual papers he/she has written. On the other hand, author-author citation relationship may not be avoided since prominent authors tend to write papers with their students or other (potentially) prominent authors. Note that we have excluded author-author social collaboration relationship from our current study due to lack of data. \textcolor{black}{Finally, we have excluded the author-author cocitation partly due to its less-intuitive nature, and mostly due to restricting computation by reducing the network dimension.} On the summary, in our current study, we use three relationships -- paper-paper citations, author-author citations and coauthor relationships among authors.

\textcolor{black}{One particular issue that may seem confusing to the reader is that here we use paper-paper citation and author-author citation relationships simultaneously as features, where one can easily derive the latter from the former.} An intuitive justification may be given as follows: a paper that is cited by a recent paper of an eminent researcher may receive equal credit to a citation by an unknown author; however by considering author-author citation we impost more credit to the former citation than the latter.

We now try to outline the proposed author ranking strategy and the underlying network model on which the proposed strategy would be applied.

\section{Network model and the outline of the  proposed ranking strategy}
\textcolor{black}{In this paper, we propose a PageRank based multi-featured author indexing strategy called \emph{$C^3$-index} (abbreviation of paper-paper {\bf C}itations, author-author {\bf C}itations and author-author {\bf C}ollaborations) that may resolve generalized opinion among the majority class of low-profile authors. We shall see shortly that the proposed ranking scheme -- is found to be consistent, effectively resolves the uncertainty among low-ranked authors, and may be used to predict future achievers in the early stage of their research career.}

\begin{figure}[!t]
 \centering
 \includegraphics[scale=.20]{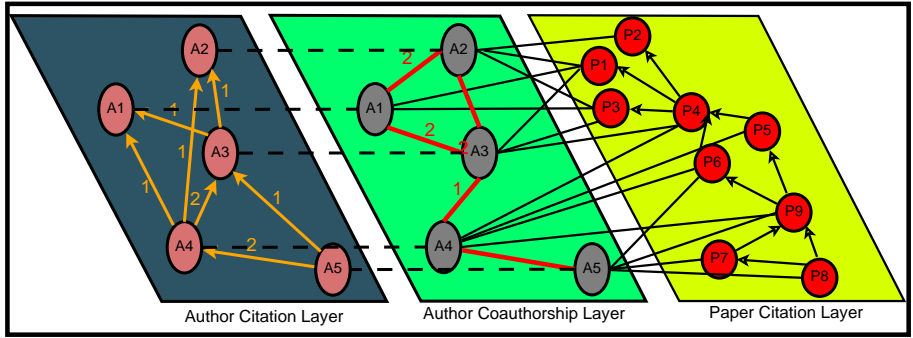}
 \caption{ \textcolor{black}{Three-layer network model} used in $C^3$-index for ranking authors. Individual layers are: (i)  Author citation layer -- a weighted directed network, where vertices are the authors, and weighted edges are drawn from vertex $A_j$ to $ A_i $ if author $ A_j $ cites the papers of author $ A_i $, the weight of the edge being the number of papers of author $ A_i $ being cited by author $ A_j $; (ii) Author coauthorship layer -- a weighted undirected network where vertices are authors, and undirected weighted edges are given between authors who jointly published papers, the weight of the edge being the number of papers the pair coauthored; (iii) Paper citation layer -- a directed network where vertices are the papers, and edges are drawn from paper $ P_j $ to paper $ P_i $, if paper $ P_j $ cites paper $ P_i $. \textcolor{black}{Lastly, there are inter-layer edges from author $ A_i $ to paper $ P_j $, if one of the authors in paper $ P_j $ is $ A_i $.}  }
\label{figure:Network_Model}
\end{figure}

\begin{figure}[!h]
\begin{minipage}{1.0\textwidth}
\renewcommand{\arraystretch}{1.2}
\centering
\scalebox{1.0}{
\begin{tabular}{|l|c|c|c|c|}
\hline
{\bf Author} & {\bf h-index} & {\bf g-index} & {\bf $ C^3 $-index} & {\bf ACI}, {\bf PCI}, {\bf AAI}\\
\hline
B. Bollobas (E) & 1 & 1 & 7.88 & 0.45, 4.68, 2.54\\

B. Shneiderman (A) & 13 & 20 & 54.86 & 23.12, 18.12, 13.42\\

G. Rozenberg (F) & 4 & 5 & 26.81 & 2.94, 14.44, 9.21\\

H. V. Jagadish (B) & 11 & 16 & 17.66 & 6.50, 5.70, 5.24\\

M. S. Hsiao (G) & 4 & 5 & 2.21 & 0.78, 0.64, 0.58\\

Ronald L. Rivest (C) & 9 & 27 & 79.02 & 39.58, 28.07, 11.17\\

S. Shelah (H) & 2 & 3 & 15.17 & 0.44, 8.29, 6.24\\

Tova Milo (D) & 7 & 11 & 6.06 & 2.26, 1.74, 1.86\\
\hline
\end{tabular}}
\end{minipage}\hfill
\begin{minipage}{1.0\textwidth}
 \centering
 \includegraphics[scale=.3]{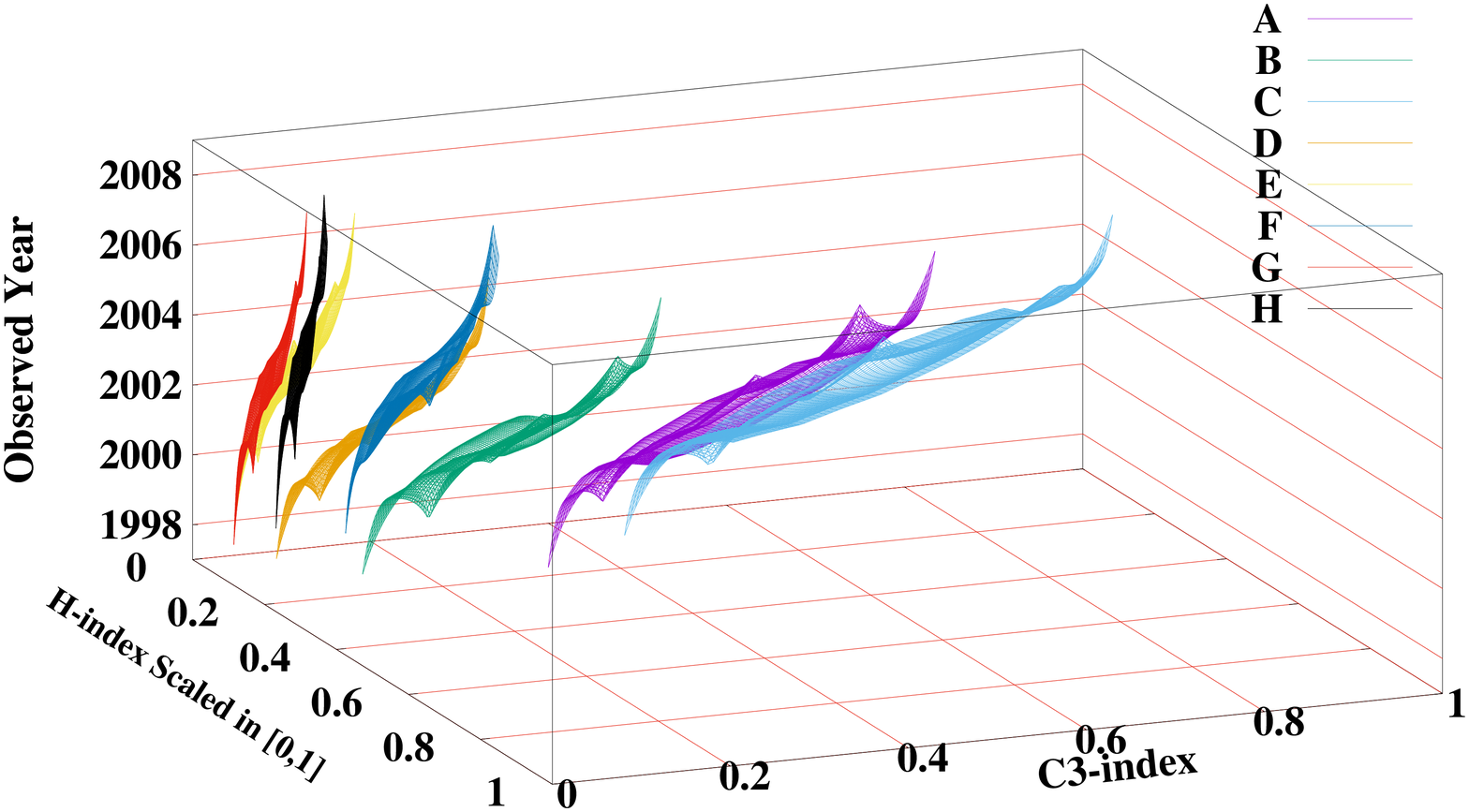}
\end{minipage}
\caption{\textcolor{black}{In the table, three component scores in $C^3$-index scoring strategy, viz. the \textbf{A}uthor \textbf{C}itation \textbf{I}ndex (ACI), \textbf{P}aper \textbf{C}itation \textbf{I}ndex (PCI), and \textbf{A}uthor co\textbf{A}uthorship \textbf{I}ndex (AAI) for eight selected authors are compared with their respective h-index and g-index (the authors are selected from the results shown in Figure \ref{figure:Scatter_Plot}, which will be discussed later). All the values of the metrics shown in the table are for the year 1998. A strong correlation may be observed between h-index, g-index and ACI component of the proposed $C^3 $-index; but the same correlation is weak correlation with the other two components. These correlations suggest that the citation-based author ranking indices like h-index and g-index may nicely capture the effect of ACI component of the proposed $ C^3 $-index, but fail to capture the effects of the other two components. The 3D plot in the figure alongside shows the changes in h-index and $C^3$-index over the years for all the selected authors mentioned in the table during the year range 1998 to 2008. To maintain clarity of the figure, the h-index values are scaled within the range [0,1] by dividing actual h-index of the corresponding author by the observed maximum h-index value for an author in the dataset. Authors having higher $C^3$-index in 1998 show steeper growth both in h-index and in $C^3$-index as they progressed over the year, which may be an indication that $C^3$-index somehow captures the future success in advance.}}
\label{Table:ECC_Component_Distribution}
\end{figure}

The $C^3$-index  is developed on an underlying multi-layered citation-collaboration network model described in Figure \ref{figure:Network_Model}, where three layers from left to right correspond respectively to author-author citation network, author-author coauthorship network, and paper-paper citation network. The desired $C^3$-index score is obtained by the sum of three individual component scores obtained from three layers, scores being normalized in such a way that the sum of scores of all the authors is unity. \textcolor{black}{The component scores from individual layers are computed using PageRank based strategies on respective layers of the network.} The strategy is elaborated in Section \ref{method}. The table in Figure \ref{Table:ECC_Component_Distribution} shows the $C^3$-index scores for eight selected authors along with individual score from each layer. For better visualization of the scores, the $C^3$-index score and its components are multiplied by the number of authors in dataset for the particular year, so that \emph{the average $C^3$-index score for a particular year is always unity}. For the sake of comparison, we compute both h-index and g-index scores for the same authors in the table.

\textcolor{black}{As we observe in Figure \ref{Table:ECC_Component_Distribution}, the surfaces corresponding to higher $C^3$-index scores in the beginning have steeper progress over the years both for h-index and $C^3$-index. This may be an indication that $C^3$-index can predict in advance the future success of the authors.} This may be due to the AAI and PCI components in the $C^3$-index score (see Section \ref{method} for detailed description), the former capturing the coauthor influence to the corresponding author, whereas the latter capturing the credit share of the authors due to their other coauthors. The rest of the paper is devoted to characterize $C^3$-index and to critically analyze whether it could be used for the purpose of predicting future prospect.

\section{Materials and Methods}\label{method}
\subsection{Dataset collection, filtering and representation} 
We crawled a massive publication dataset related to Computer Science domain from Microsoft Academic Search (MAS), one of the largest archived datasets. Crawling of the Microsoft Academic Search  started in October, 2015. The automated crawler initially used the ranklist given by MAS to obtain the list of paper IDs. The paper IDs were then used to fetch the metadata of the publications. We used Tor to distribute our crawling to different systems in order to avoid overloading a particular server with bursty traffic. We employed random exponential back-off time whenever the server or the connection returned some error and sent the request again. We followed the robot restrictions imposed by the servers to ensure efficient crawling of data from both client and server perspective. It took us around 6 weeks to completely crawl all the information related to the 7 million papers \cite{ChakrabortySTGM13,0002SGM14}.

The crawled data had several inconsistencies that were removed through a series of steps.  We filtered out all such papers that did not have the bibliographic attributes required for our study such as the unique index of the paper, the year of publication, the list of authors, the publication venue. We also removed few forward citations which pointed
to the papers published after the publication of the source paper. \textcolor{black}{Further, we considered only those papers published in between 1950 and 2012, that cite or are cited by at least one paper (i.e., we removed disconnected nodes with zero in-degree and zero out-degree).} The filtered dataset contains around $6$ million papers. Some of the references that pointed to papers absent in our dataset (i.e., dangling references) were also removed from the dataset. Some general information pertaining to the dataset are shown in Table \ref{general}.

\begin{table}[h!]
 \begin{center}
 \caption{General information of raw and filtered datasets.}\label{general}
\scalebox{1.1}{
\begin{tabular}{|l|r|r|}
\hline
  & Raw & Filtered \\\hline
Number of valid papers & 7,473,171 & 6,643,906\\\hline\hline
Number of papers with no venue & 343,090 & -- \\\hline
Number of papers with no author & 45,551 & -- \\\hline
Number of papers with no publication year & 191,864 & -- \\\hline
Number of authors & 4,186,412 & 3,186,412 \\\hline
Avg. number of papers per author & 5.18 & 5.04 \\\hline
Avg. number of authors per paper & 2.49 & 2.67 \\\hline
Number of unique publication venues & 6,143 & 5,938 \\\hline\hline
\textcolor{black}{Number of paper-paper citation edges} & -- & \textcolor{black}{54,794,224} \\\hline
\textcolor{black}{Number of coauthorship edges} & -- & \textcolor{black}{10,837,179} \\\hline
\textcolor{black}{Sum of weights - coauthorship edges} & -- & \textcolor{black}{19,718,437} \\\hline
\textcolor{black}{Number of author-citation edges (excluding self loops)} & -- & \textcolor{black}{176,174,616} \\\hline
\textcolor{black}{Sum of weights - author citation edges} & -- & \textcolor{black}{371,572,974} \\\hline

\end{tabular}}
\end{center}
\end{table}

\if{0}
In our present study, we used the dataset developed by Tang et al. ~\cite{Arnetminer_Dataset} which is available freely in Arnetminer Project. The original dataset had 2,244,021 papers and 4,354,534 citation relationships. We observed some inconsistencies in the dataset in the form of one or more missing data fields. Those data elements that has missing fields (except the abstract field) is not usable in our current study and hence are filtered out from the dataset. One particular information component missing in the above dataset is the field/domain of research a particular paper may be associated, which is an important consideration in our present work. Dataset prepared by Chakraborty et al~\cite{Tanmoy_ASONAM_2013} is used for finding the field of interest of research for a particular author and augment the same for the authors in our dataset.\\
\fi

\subsection{Network construction} From the filtered dataset, we prepare the multilayer network. The creation of the paper-paper citation network is easy -- we \textcolor{black}{consider} the papers as the vertices, and connect vertex $ P_i $ to vertex $ P_j $, if the paper corresponding to $ P_i $ cites the paper corresponding to paper $ P_j $. For preparing the other two layers, we need to extract the author information. To do that, the names of authors are extracted from the author lists for all the papers. \textcolor{black}{To remove the ambiguity from author name, we use ``RankMatch'' algorithm proposed by Liu et al. \cite{Liu:2013}.} There are a couple of reasons behind adopting this algorithm. First of all, it is a completely unsupervised approach which is required in our study. In addition, the algorithm has been proved to be effective for the same types of scientific dataset. The algorithm first assigns a unique index ID to all the author names present in the dataset. Then it follows a two-step strategy -- (i) For each indexing author ID, it tries to pull out all the authors whose names are possible variations of the indexing author name. To come up with the pool, it takes into account a number of cases where names can mutate or be disturbed. (ii) In the second step, it trims the candidate pool based on authors' publication features. Examples of publication features include  publication venues, years, and title words. These features turn out to be discriminative for identifying real duplicates from the candidate pool. Once the unique authors are extracted, they are given suitable Author Identifiers for further references. For each author in the Author List obtained here, we add one vertex each in the coauthorship layer as well as in the author-author citation layer. For each paper $ P $ in the dataset, an edge between vertex $ A_i $ and vertex $ A_j $ is added in author-author coauthorship network, if both the authors corresponding to the IDs are in the author list of paper $ P $. If there exists another paper $p'$ for which $ A_i $ and $ A_j $ coauthored together, the weight of the edge between vertex $ A_i $ and vertex $ A_j $ is incremented by one. For preparing the author-author citation network, we check if paper $ P_i $ cites paper $ P_j $, then all the authors of $ P_i $ have links to all the authors of $ P_j $. If some pair of authors already has links between them, its weight is incremented by one. {\color{black} Note that we also remove self-citation in the construction of the author-author citation network. We consider those citations as ``self-citations'' where at least one author is common in both citing and cited papers as defined in \cite{Carley2013}.} Once the network is created, the iterative modified PageRank algorithm discussed below are executed and the values for each vertex from different layers are collected.

\subsection{Measuring \emph{$C^3$}-index} 
The proposed $C^3$-index is computed using a set of iterative formulas. The $C^3$-index of the $ j^{th} $ author $ A_{j} $ at iteration level $ t $, denoted by $ C_j^{3(t)} $, is obtained as:
			\begin{equation*}
				\begin{split}
					C_j^{3(t)} = (1-\theta) + \theta \times (ACI_j^{(t)}\ +\ AAI_j^{(t)}\ +\ PCI_j^{(t)})
				\end{split}
			\end{equation*}

In the above formula the terms 	$ ACI_j^{(t)} $ and $ AAI_j^{(t)} $, denote the scores of author $A_j$ in author-author citation network and the author-author coauthorship network, respectively, that are obtained using the following iterative formulas:
			\begin{equation*}
				\begin{split}
					ACI_j^{(t)}  = (1-\theta) + \theta \times \sum_{A_k\:\in\:C(A_j)}\dfrac{ACI_k^{(t-1)}}{outdeg(A_k)}
				\end{split}
			\end{equation*}

    		\begin{equation*}
				\begin{split}
					AAI_j^{(t)}  = \sum_{A_k\:\in\:CA(A_j)} \dfrac{AAI_k^{(t-1)}}{deg(A_k)}
				\end{split}
			\end{equation*}

\noindent where $ C(A_j) $ denote the set of authors who cited at least one paper of author $ A_j $, $ CA(A_j) $ denote the set of authors who coauthored with author $ A_j $ in at least one paper, $ outdeg(A_k) $ denotes the sum of the degrees of the outgoing edges from node $ A_k $ in the author-author citation layer of the network, $ deg(A_k) $ denotes the sum of the degrees of the edges incident on node $ A_k $ in the author coauthorship layer, and $ \theta $ is the \textit{damping factor} for the PageRank based strategy. In our experiments, it is set to $0.5$ following the suggestion made by Chen et al.~\cite{finding_scientific_gems_using_google_pagerank}.

The third component in the formula, $ PCI_j^{(t)} $ denotes the paper citation index score for author $A_j$ at the iteration level $ t $ that are obtained from the paper citation layer of the network. It is the sum of the paper credits shared at that level for the publications made by author $ A_{j} $ distributed uniformly (or some other rule) among all the coauthors of the paper using the formula:	
			\begin{equation*}
				\begin{split}
					PCI_j^{(t)}  = \biggl(C_j^{3(t-1)}\biggl)^{\alpha} \times \sum_{P_k\:\in\ P(A_j)\:} 												\dfrac{PQI_k^{(t-1)}}{\sum_{A_l\:\in\ A(P_k)\:} \biggl(C_l^{3(t-1)}\biggl)^{\alpha}} 
				\end{split}
			\end{equation*}

\noindent where $ P(A_j) $ denote the set of papers published by the author $ A_j $, $ A(P_k) $ denote the set of authors for the paper $ P_k $, and $ PQI_k^{(t)} $ is a paper quality index score representing the credit of the paper that is obtained from the paper citation layer of the network using a PageRank based algorithm as follows:
			\begin{equation*}
				\begin{split}
					PQI_i^{(t)}  = (1-\theta) + \theta \times \sum_{P_k\:\in\:C(P_i)} \dfrac{PQI_k^{(t-1)}}{outdeg(P_k)}
				\end{split}
			\end{equation*}
			
\noindent where $ C(P_i) $ denote the set of papers citing paper $ P_i $, and $ outdeg(P_k) $ denote the number of the outgoing edges from node $ P_k $ of the paper citation layer. We use the same damping factor $ \theta $ for all the PageRank formulas mentioned here.

As a final note, we represent $ PCI_j $ as a generalized formula, where $ \alpha $ is used as a \emph{model parameter} to decide the way credit from an individual paper would be distributed among its authors. If it is set to 0, as is the case in the current experiments, then the credit will be distributed uniformly to all the coauthors. But for other values of $ \alpha $, the credit will be distributed on the basis of their current $ C^3 $-index. If $ \alpha $ is positive value, then authors having higher $ C^3 $-index would receive larger share of the credit, whereas if $ \alpha $ is negative, the authors with lower $ C^3 $-index would receive larger share.

\section{Results}

\subsection{$C^3$-index vs. H-index}
An immediate question would be how the ranking produced by h-index differs from the ranking obtained from $C^3$-index and its individual components. To verify this, we measure the Spearman Rank Correlation Coefficient between the pair-wise ranks (Table \ref{Table:Correlation}). The coefficient values suggest a strong correlation of h-index with ACI component, but relatively narrow correlation with the other two. This once again corroborate with our earlier observation in Figure \ref{Table:ECC_Component_Distribution}.

\begin{table*}[!h]
\renewcommand{\arraystretch}{1.3}
\centering
\caption{The Spearman Rank Correlation Coefficient between h-index, $C^3$-index and its components. The values indicate that h-index is highly correlated with the ACI score, as compared to that for other two components, and hence with $C^3$-index as a whole. Thus we hypothesize that the information carried by $C^3$-index would be significantly different from that of h-index.}
\label{Table:Correlation}\label{table:corr}
\begin{tabular}{|c|c|c|c|c|}
\hline
Year & H-index vs $C^3$-index & H-index vs ACI & H-index vs PCI & H-index vs AAI \\
\hline
1998 & 0.577136 & 0.989151 & 0.467660 & 0.401122 \\
\hline
2004 & 0.604968 & 0.988483 & 0.517128 & 0.426008 \\
\hline
2008 & 0.613174 & 0.988427 & 0.539801 & 0.437871 \\
\hline
\end{tabular}
\end{table*}

\begin{figure}[!h]
\centering
  \subfigure[]{
  \includegraphics[scale=.42]{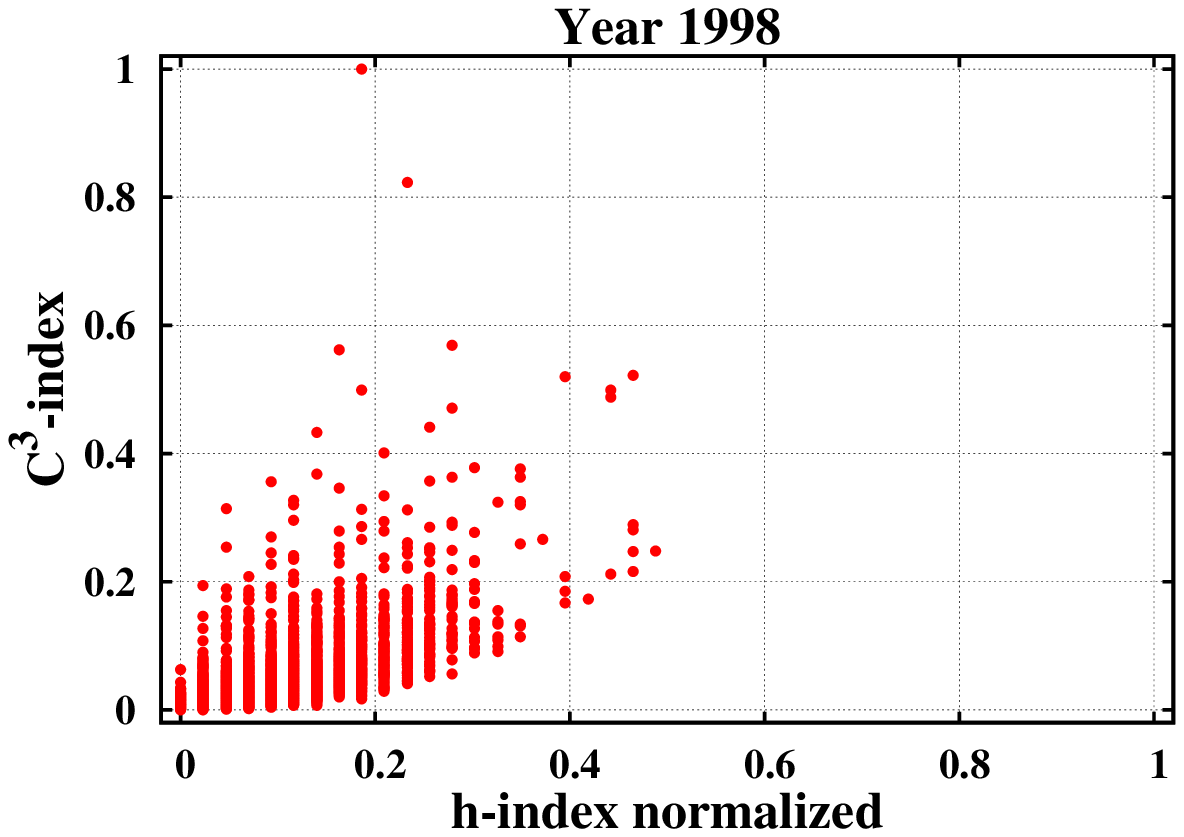}
  \label{figure:Scatterplot_1998_h}
   }
  \subfigure[]{ 
  \includegraphics[scale=.42]{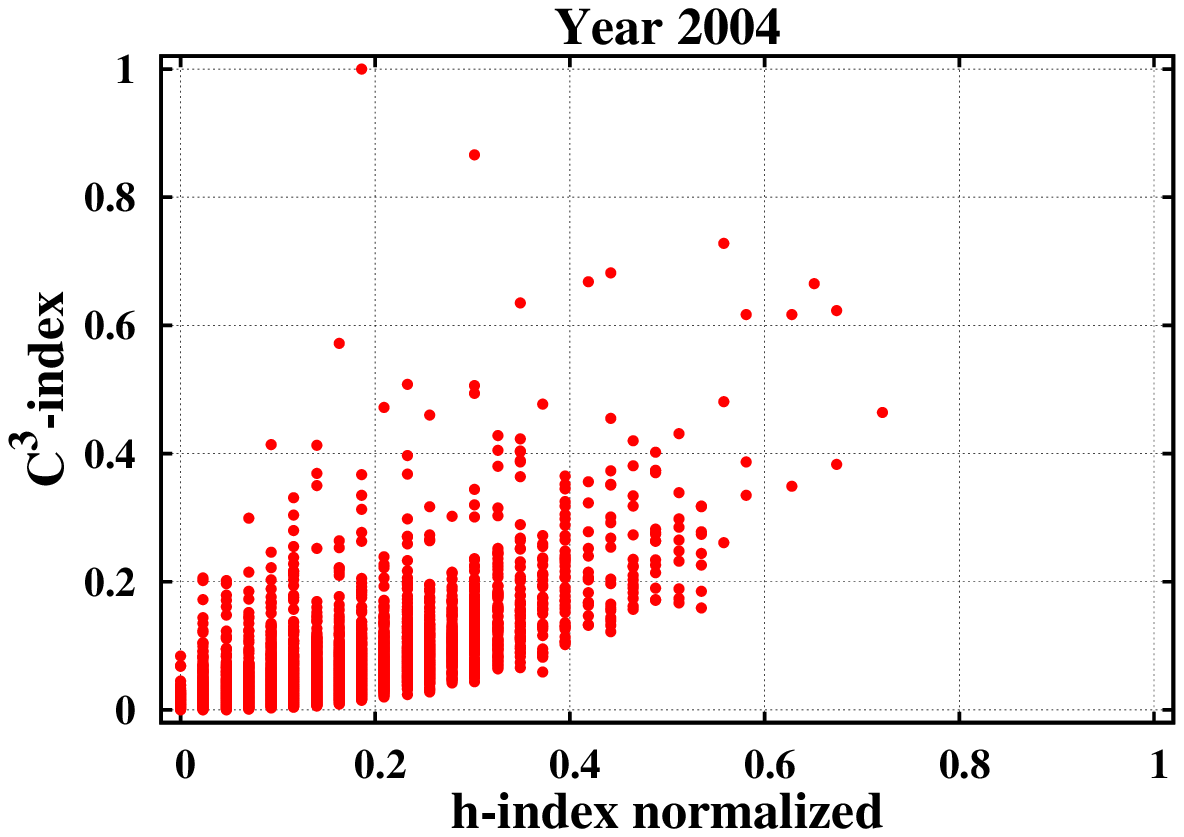}
  \label{figure:Scatterplot_2004_h}
   }
  \subfigure[]{ 
  \includegraphics[scale=.42]{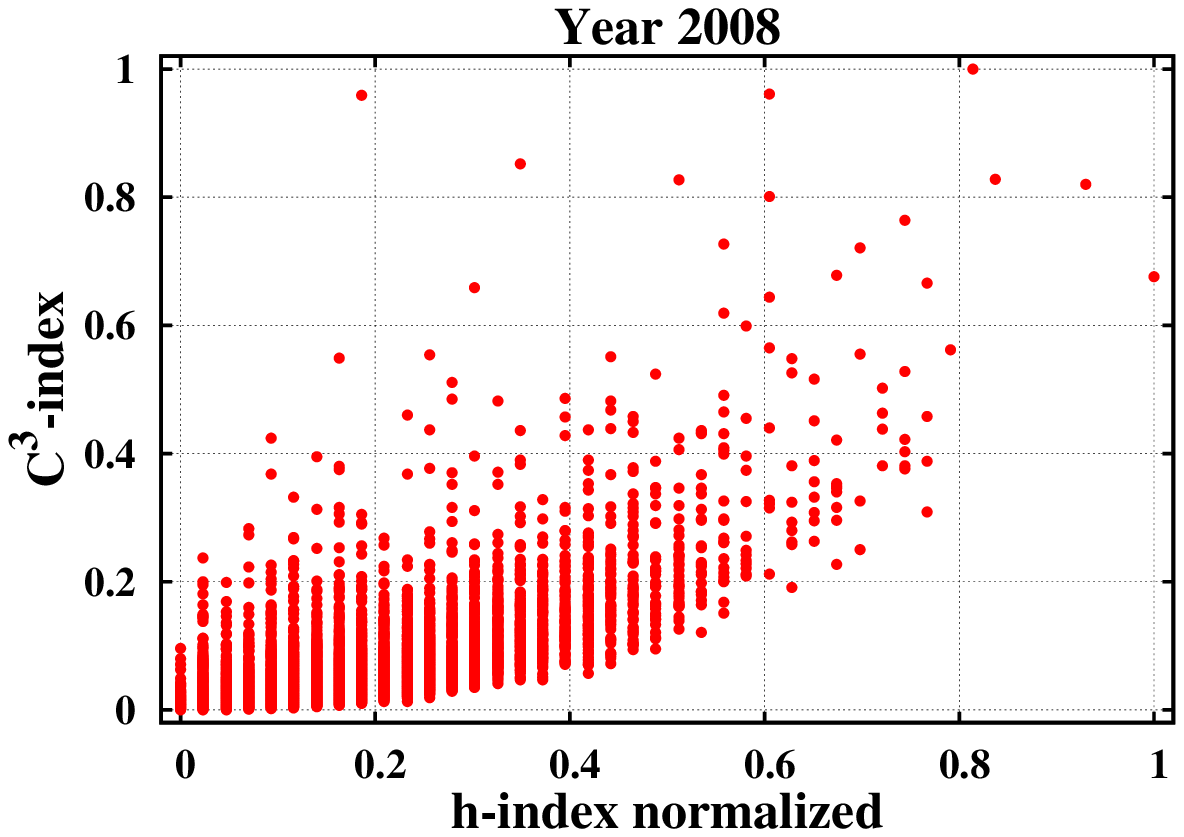}
  \label{figure:Scatterplot_2008_h}
   }
  \subfigure[]{
  \includegraphics[scale=.42]{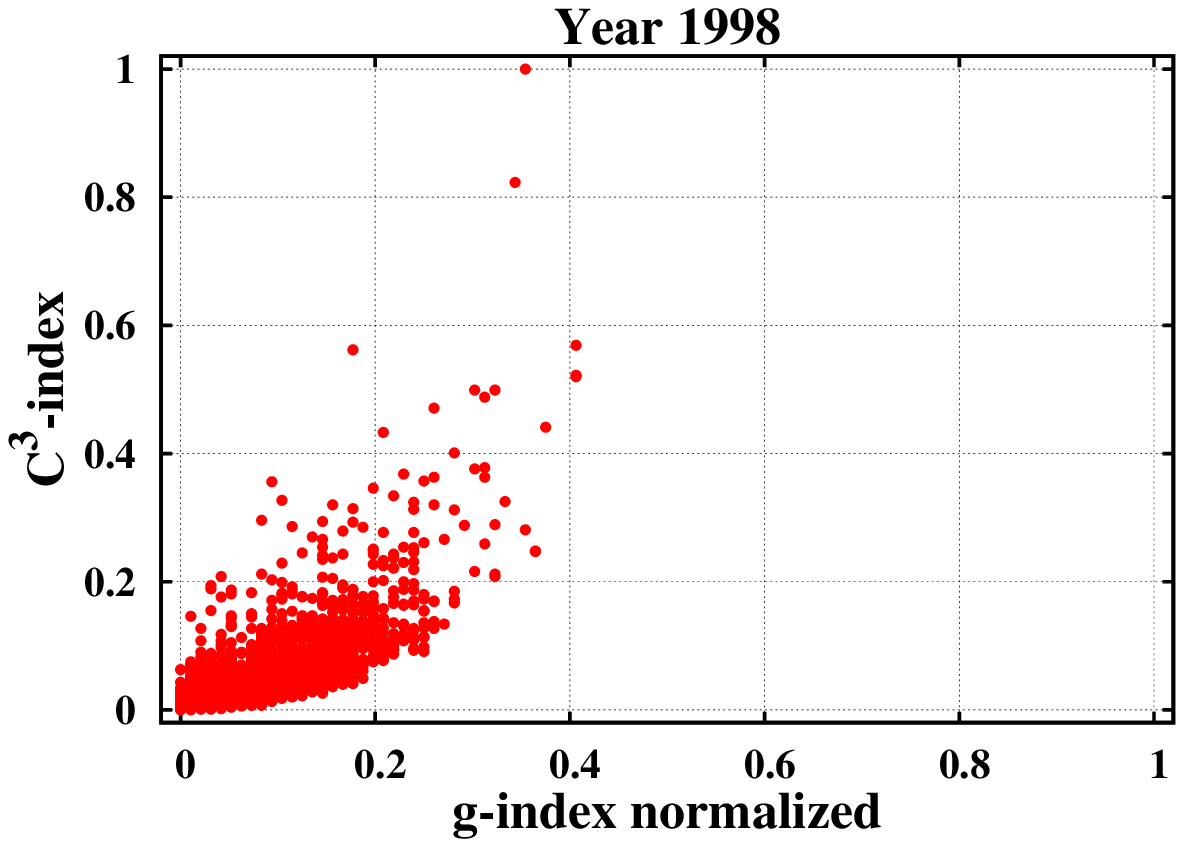}
  \label{figure:Scatterplot_1998_g}
   }
  \subfigure[]{
  \includegraphics[scale=.42]{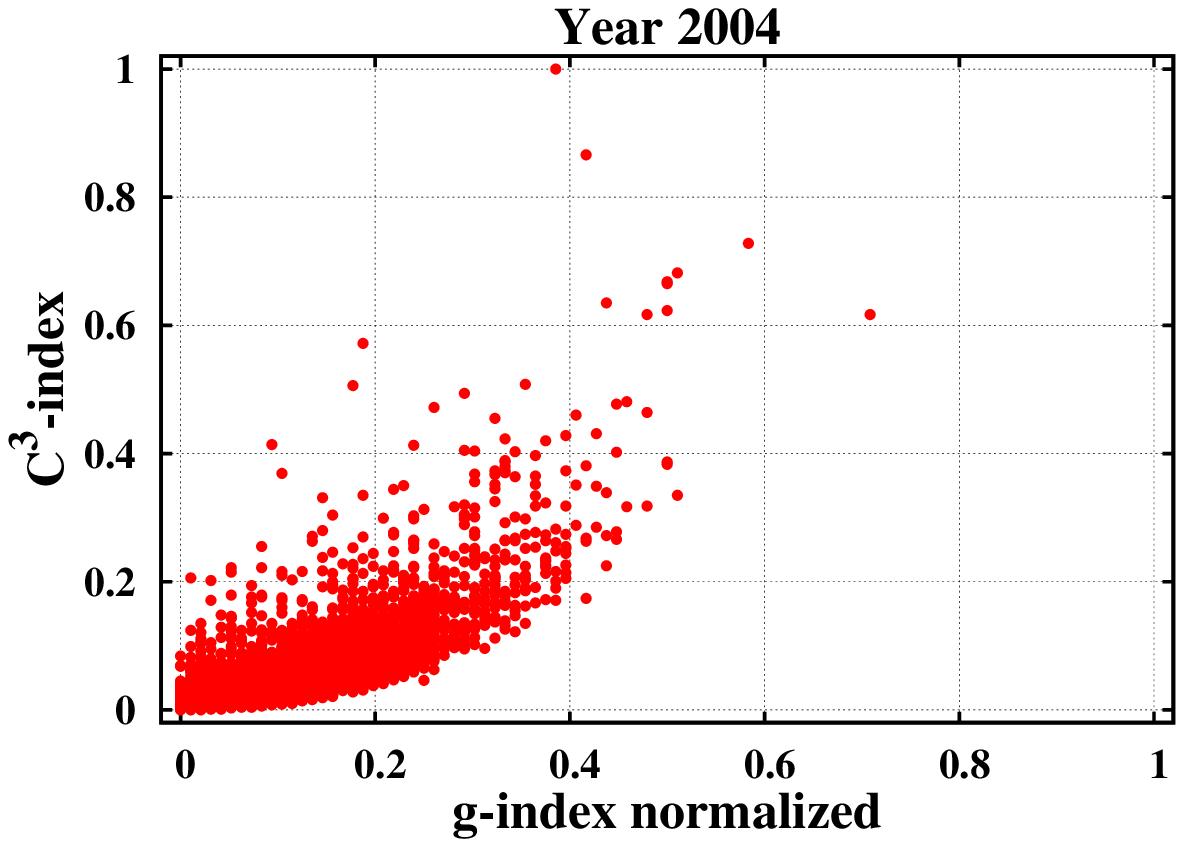}
  \label{figure:Scatterplot_2004_g}
   }
  \subfigure[]{
  \includegraphics[scale=.42]{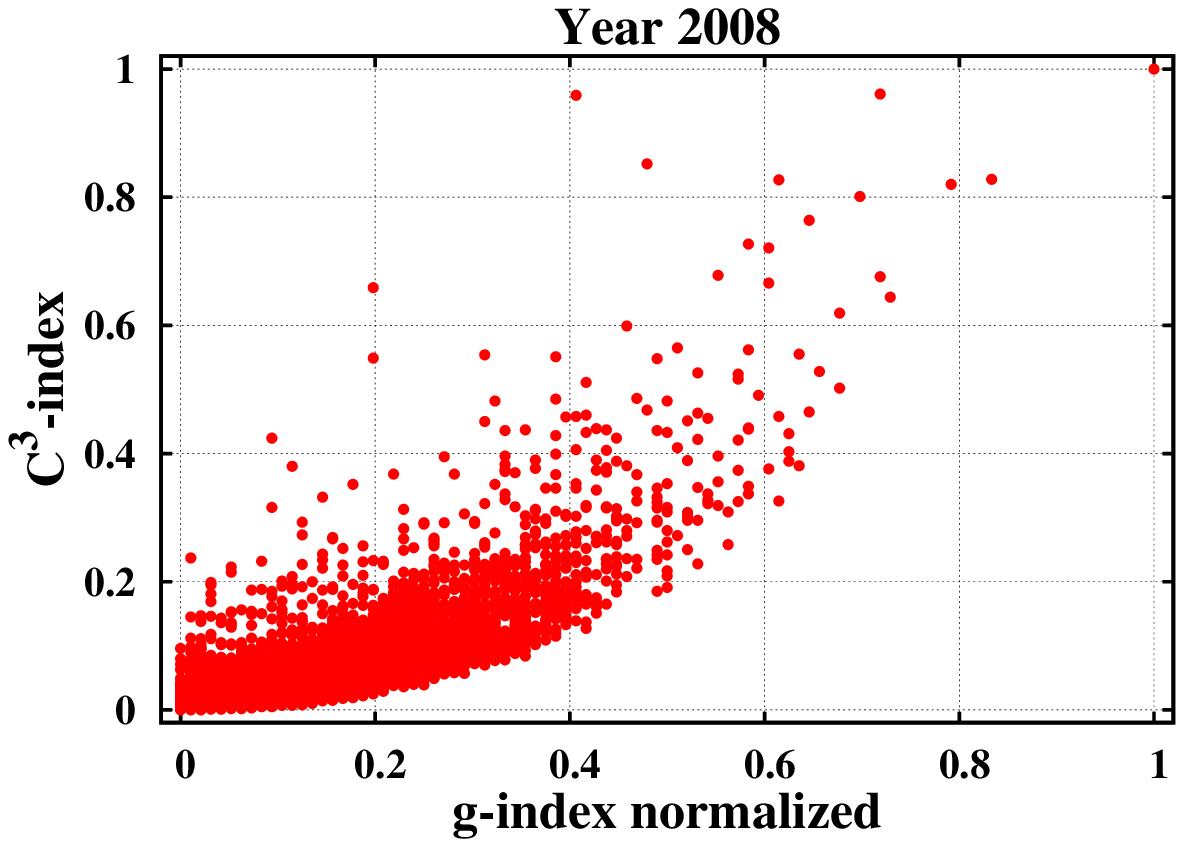}
  \label{figure:Scatterplot_2008_g}
   }
\caption{Scatter plots in the figure show distribution of $C^3$-index against h-index (top panel) as well as against g-index (bottom panel) for all the authors in the dataset during the years 1998, 2004 and 2008. Both h-index and g-index are scaled down within the range of 0 to 1 by dividing the actual index values by the highest value of the corresponding index in the time period of 1998-2008. In all the figures, we observe that the value of $C^3$-index for majority of the authors remains almost consistent with their respective h-index as well as with their g-index. However, we observe few inconsistent points mostly in upper-left portion of the plots, indicating those authors having low h-index (g-index), but high $C^3$-index. This is possibly an indication of low citation but high coauthorship credit for the corresponding authors. In Figure \ref{Table:ECC_Component_Distribution}, we selected some of authors having such inconsistencies and analyzed their behavior over the years.}
\label{figure:Scatter_Plot}
\end{figure}

In Figure \ref{figure:Scatter_Plot}, we show the correlation between $C^3$-index and h-index for all the authors in the dataset in a different manner. In all the sub-plots in Figure \ref{figure:Scatter_Plot}, we plot the author scores obtained using $C^3$-indexing strategy for all the authors in the dataset against their respective h-index and g-index. The $C^3$-index as well as the h-index and g-index are calculated for a particular year by considering the publication entries in the dataset up to that particular year (i.e., by removing from the dataset the papers which are published after that year, the citations that are made after that year, and the authors who made their first publication after that year). The same procedure is followed for all the temporal studies made in this paper. In other words, as the year of study proceeds towards the current time, the data volume increases gradually in all respect. 

In Figure \ref{figure:Scatter_Plot}, the points close to the diagonal of each subplot represents those authors whose $C^3$-index values are perfectly correlated with h-index (g-index) strategy. However, we further observe that there are few authors with low h-index but high $C^3$-index (upper-left portion), and vice versa (lower-right portion). We selected some of these authors earlier and analyzed the profiles in Figure \ref{Table:ECC_Component_Distribution}.


\subsection{Temporal growth pattern}
\begin{figure}[!h]
\begin{minipage}{0.55\textwidth}
\centering
  \subfigure[]{
  \includegraphics[scale=.21]{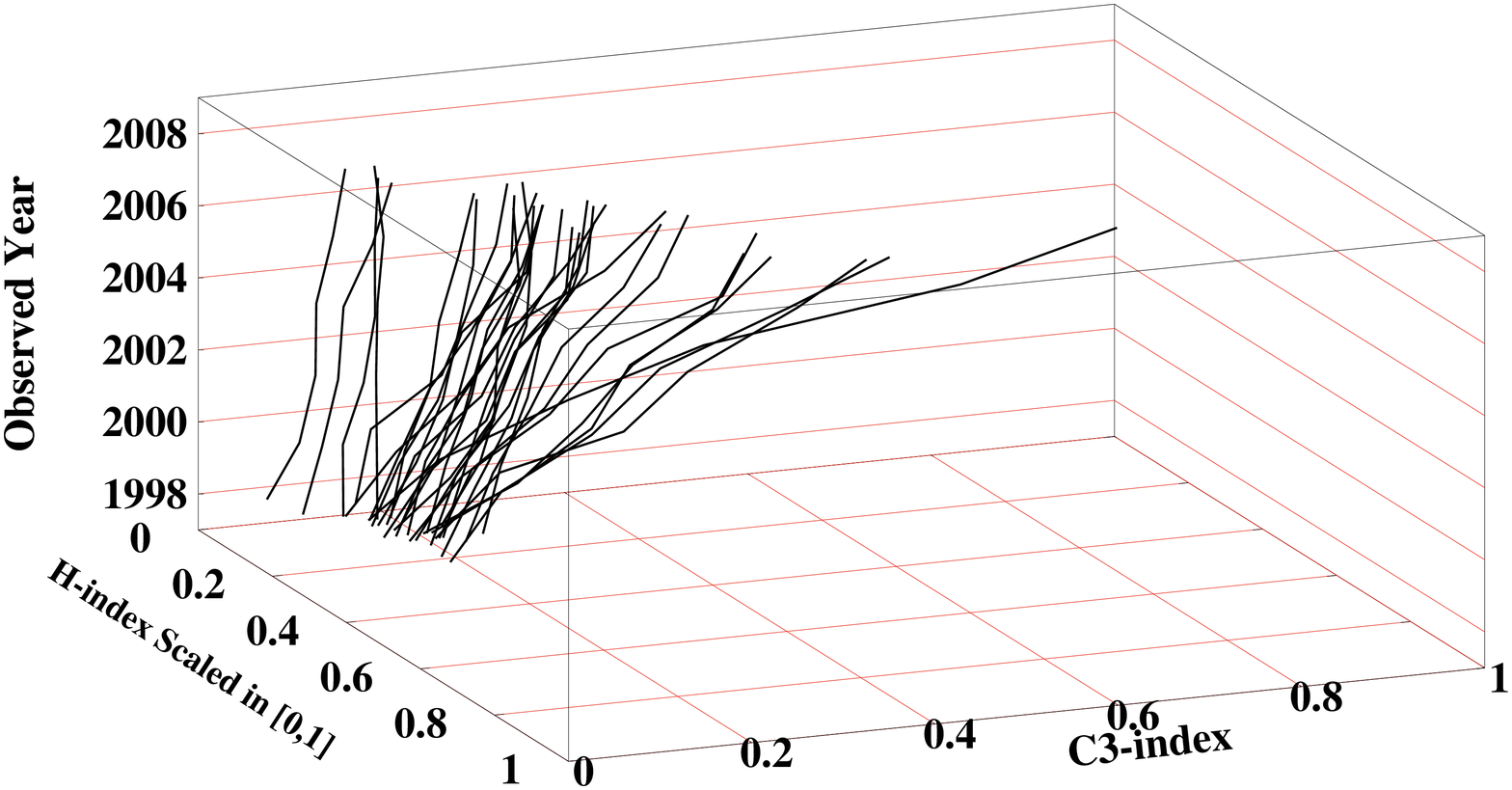}
  \label{figure:Low_ACI_High_AAI}
   }
  \subfigure[]{ 
  \includegraphics[scale=.21]{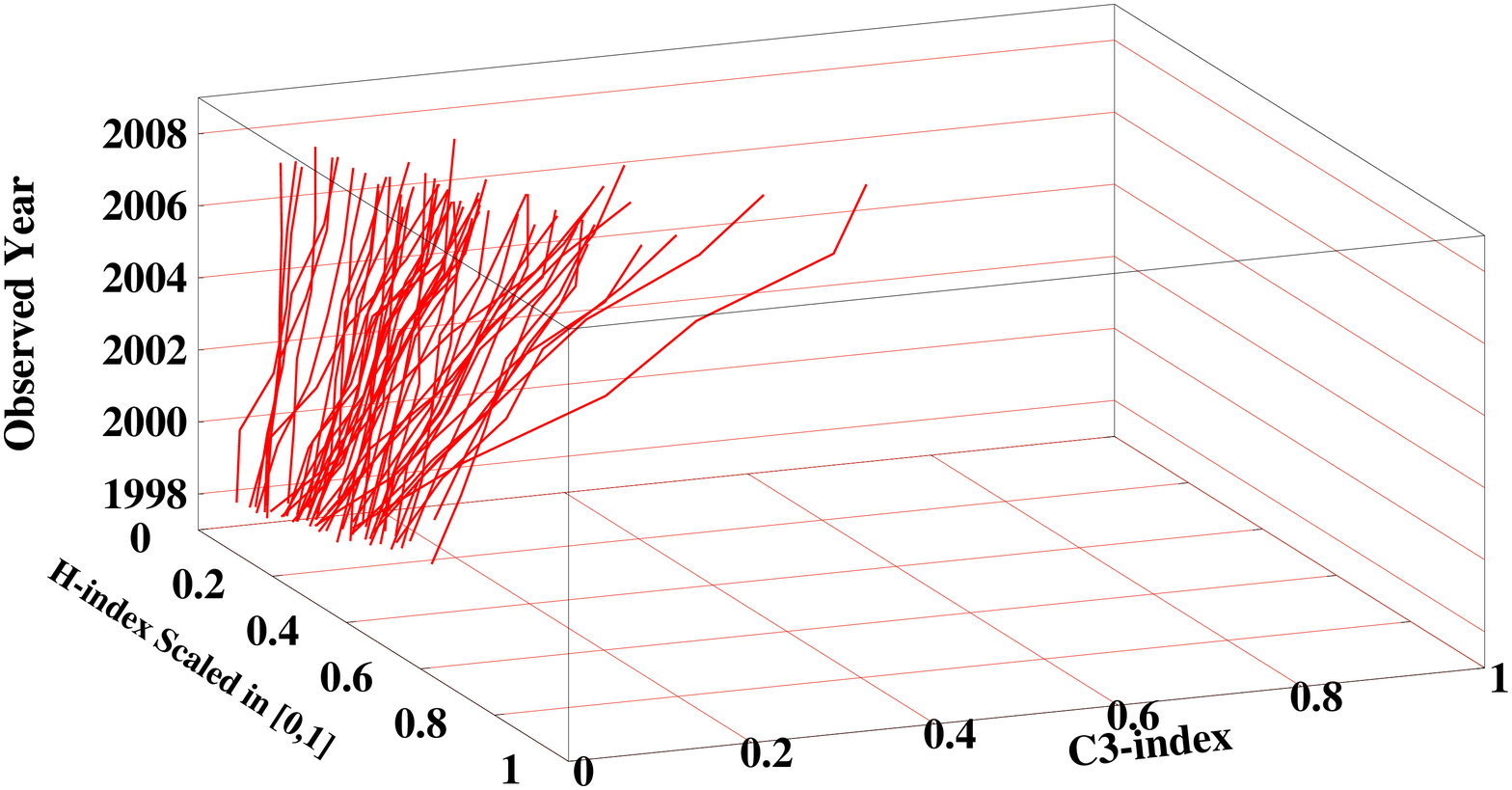}
  \label{figure:Low_ACI_Low_AAI}
   }
\end{minipage}\hfill
\begin{minipage}{.45\textwidth}
 \centering
 \subfigure[]{ 
  \includegraphics[scale=.21]{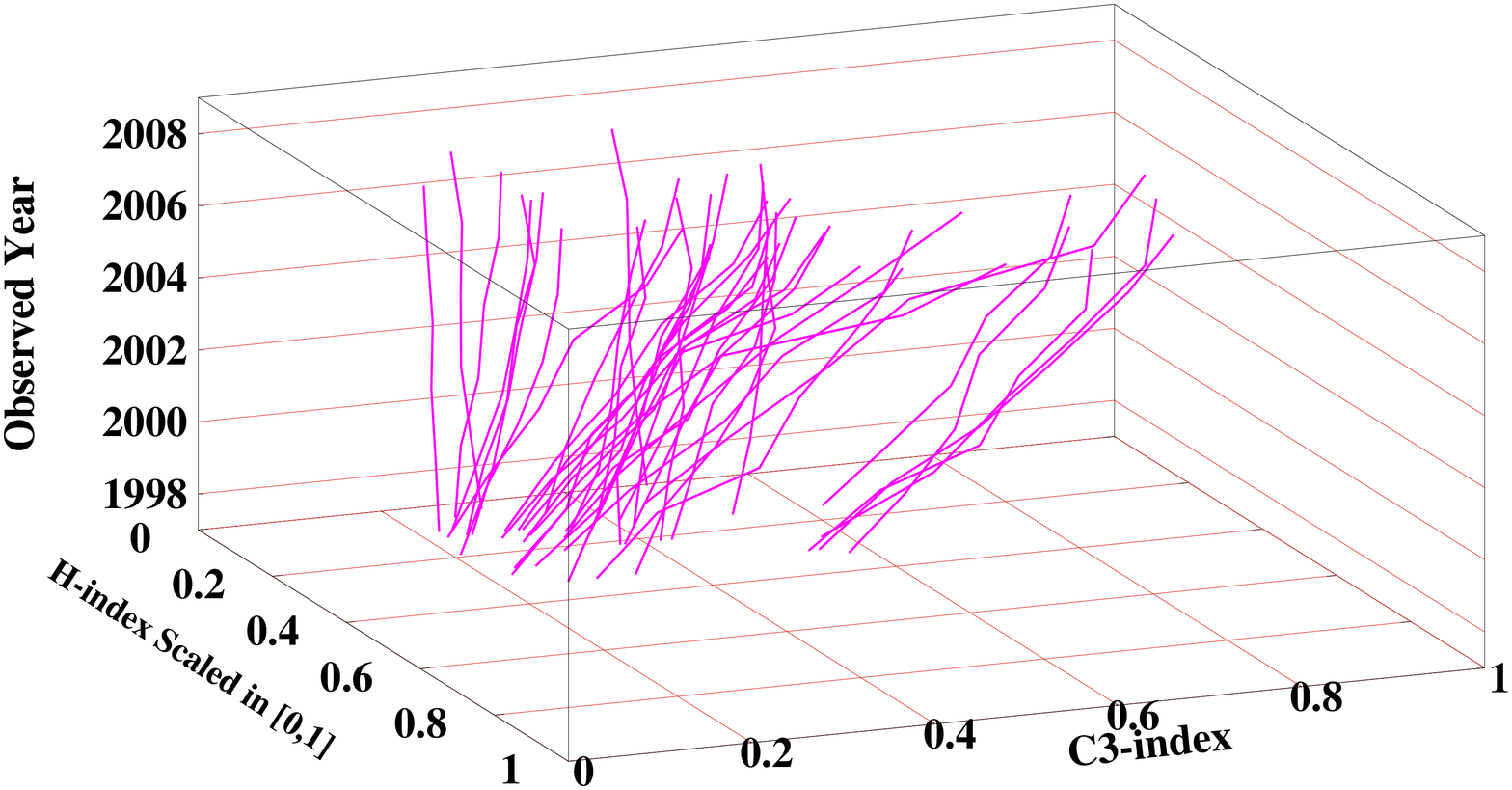}
  \label{figure:High_ACI_High_AAI}
   }
   \subfigure[]{ b
  \includegraphics[scale=.21]{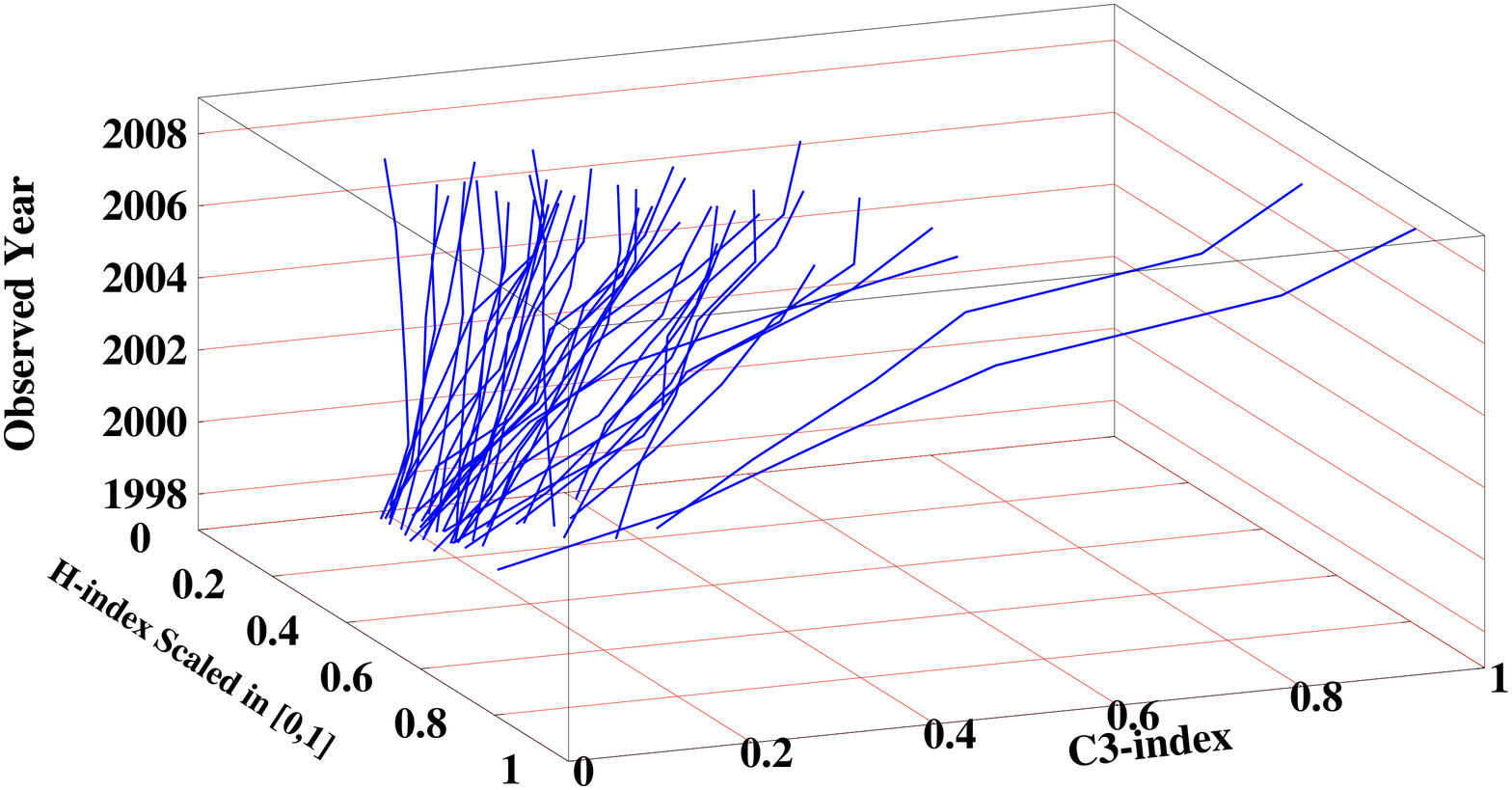}
  \label{figure:High_ACI_Low_AAI}
   }
\end{minipage}
\caption{\textcolor{black}{Proposed $C^3$-index has three components: ACI, PCI and AAI, respectively. We observed that h-index (and also g-index) has high correlation with ACI component, but has low correlation with the other two (Table \ref{table:corr}). Here we select four sets of authors: (a) authors having ACI $\leq$ 20\% of the $ ACI_{max} $, AAI $\geq$ 80\% of $ AAI_{max} $, (b) authors having ACI $\leq$ 20\% of the $ ACI_{max} $, AAI $\leq$ 20\% of $ AAI_{max} $, (c) authors having ACI $\geq$ 80\% of the $ ACI_{max} $, AAI $\geq$ 80\% of $ AAI_{max} $, (d) authors having ACI $\geq$ 80\% of the $ ACI_{max} $, AAI $\leq$ 20\% of $ AAI_{max} $. The scores are selected on the basis of the year 1998. We plot 3D line curves for the corresponding authors in the respective sub-figures. In general, the figures suggest that the authors having high AAI-score improved more during the time period 1998-2008 than those having low AAI-scores. This suggests that the inclusion of AAI-score in the proposed $C^3$-index has brought future prediction capability in it.}}
\label{figure:Growth Gradient}
\end{figure}

In Figure \ref{figure:Growth Gradient} we study the year-wise transformation of performance indices (h-index as well as $C^3$-index) for four sets of authors selected from the authors in 1998. Figure \ref{figure:Low_ACI_High_AAI} corresponds to the set of authors who have relatively low ACI-score, but high AAI-score in 1998. We select 31 authors from this category and show their growth of the said indices over the years. In Figure \ref{figure:Low_ACI_Low_AAI} we plot similar results for 48 authors from the author pool who have low ACI-score and low AAI-score in 1998. While comparing the above two plots, we observe that the indices for most of the authors in Figure \ref{figure:Low_ACI_High_AAI} tend to end up with much higher values as compared to that for authors in  Figure \ref{figure:Low_ACI_Low_AAI}, in both the cases the lines start nearly from the same point.  This perhaps hints upon a point that the ACI component of $C^3$-index has some kind of correlation with future performance behaviour of the concerned researcher. In Figure \ref{figure:High_ACI_High_AAI} and Figure \ref{figure:High_ACI_Low_AAI} we plot similar growth curves for another two sets of authors, both having high ACI scores but different AAI scores. Here also we observe that the major portion  of authors from the author set having higher AAI scores end up with higher performance indices.

\subsection{Capturing future performance through $C^3$-index}
\begin{figure}[!t]
\centering 
 \subfigure[]{
  \includegraphics[scale=.38]{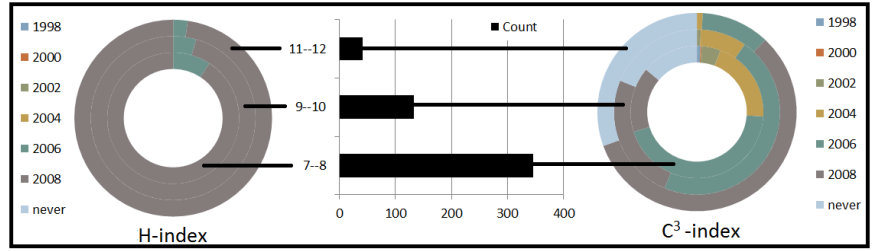}
  \label{figure:Future_low_performers}
   }
  \subfigure[]{
  \includegraphics[scale=.38]{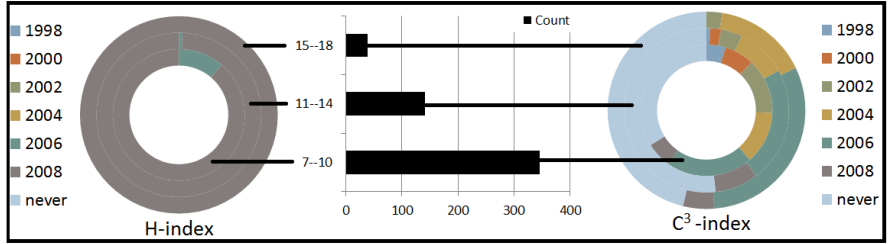}
  \label{figure:Future_Medium_performers}
  }
\caption{(a) A set of authors is extracted from the dataset having zero h-index in 1998, but acquired moderate h-index (ranging 7-12) in 2008. The bar plots in the middle show the number of those authors in three equal-sized h-index bins. \textcolor{black}{On the left-hand multi-level pie-chart, the three concentric rings correspond to the set of authors lie in the respective h-index bin associated.} Each individual ring corresponds to a pie-chart that shows a distribution over the years of authors from author subset corresponding to the associated h-index bin meeting the h-index ceiling for that bin. \textcolor{black}{The multi-level pie-chart on the right shows similar kind of distribution for the same sets of authors using their $C^3$-index over time. The $C^3$-index ranges for three concentric rings and are set to 0.02 - 0.029, 0.03 - 0.039, 0.04 - \dots . The individual rings in pie-charts represent similar author distribution over the years as on the left-hand side figure. The pie-chart on the right side suggests that in case of $C^3$-index, the change in the author score is visible much ahead of time than in the case of h-index, which indicates that the proposed strategy can capture the authors' future performance much ahead of time than h-index. (b) In order to verify whether the above observation is valid even for the cases of authors who already reached the level of medium/top rankers, a set of authors is extracted from the dataset having h-index ranging from $4$ to $7$ in 1998, but acquired moderate to high h-index (ranging 7-18) in 2008.} The bar plots in the middle show the number of those authors in three equally-divided h-index bins as shown in the figure. We plot the same multi-level pie-chart pairs similar to Figure (a). In case of right-hand side chart, the $C^3$-index bins are set to be the following: 0.08 - 0.14, 0.141 - 0.17, 0.171 - \dots. The distributions of authors in both the multi-level pie-charts suggest that proposed $C^3$-index strategy can capture the future performance much ahead of time.}
\label{figure:Multilevel_Pie_Chart}
\vspace{-5mm}
\end{figure}

In Table \ref{Table:Correlation} we already observed that h-index  has strong correlation with the ACI component of $C^3$-index, but has weak correlation with the other two components. \textcolor{black}{In Figure \ref{figure:Growth Gradient}, we intend to find whether that correlation behavior brings some meaningful insights about $C^3$-index.} The figures suggest that authors having high AAI score show rapid growth over time than those the authors with low AAI score. From this, we hypothesize that the presence of this component in $C^3$-index may provide indication of future success, which h-index and its variants perhaps lack. \textcolor{black}{To validate this, we present multi-level pie-charts in Figure \ref{figure:Multilevel_Pie_Chart} for a selected set of authors to show whether $C^3$-index is capable of predicting future success of authors in the early stage of their career.}

In Figure \ref{figure:Future_low_performers}, the set of authors who had h-index of zero in 1998, but acquired moderate h-index (ranging from $7$ to $12$) in 2008 are selected. The bar plots in the middle show the number of such authors in three equally-divided h-index bins. The multi-level pie-chart in the left shows the gradual improvement of h-index as observed over time for the authors in each bin during the time span observed in two-year separations. The multi-level pie-chart in the right points to the fraction of authors  present in respective bins shown in the bar plot exceeding a chosen $C^3$-index bound in a given year. Three different bounds are chosen for three different bins, viz. 0.02 for 7-8 bin, 0.03 for 9-10 bin, and 0.04 for 11-12 bin. In left-hand pie-chart, we pin-point the fraction of authors that reached the next h-index bin in the respective year. We observe from the left-hand pie chart that no fraction of authors reach the next h-index bin prior to 2006. On the other hand, it is apparent from the right-hand pie-chart that significant fraction of authors reach the next bin level much earlier than the above, which suggests that $C^3$-index is able to capture the change much ahead of time than h-index. This in turn establishes the predictive power of $C^3$-index.

We are now interested to see whether above mentioned future-predictive behavior of $C^3$-index holds for authors present in other portion of the author spectrum. In Figure \ref{figure:Future_Medium_performers}, a set of authors are selected whose h-index lay in the range of 4-7 in 1998. We may decently assume that such authors may be considered as medium-performers during the time when our observation begins. We observe that by 2008, the values of the selected authors' h-index lie in the range 7-18, which may indicate that some portion of the author gained high visibility (i.e., gained high h-index) in 2008; whereas the rest fail to acquire enough visibility. The bar plot in the middle shows the number of those authors in three distinct bins similar to Figure \ref{figure:Future_low_performers}. \textcolor{black}{The multi-level pie-chart in the right pinpoints the fraction of authors lying in respective bins shown in the bar plot surpassing a chosen $C^3$-index bound in a given year.} Three different bounds are chosen for three different bins, viz. 0.08 for 7-10 bin, 0.14 for 11-14 bin, and 0.17 for 15-18 bin. In left-hand pie-chart, we show the fraction of authors that reach the next h-index bin in the respective year. We observe from the diagram that major fraction of authors reach the next level after 2006, and only a small fraction reaches this level during 2006, and none does the same before 2006. \textcolor{black}{On the other hand, for $C^3$-index, future stars (those falling in 15-18 bin), are capable of surpassing the predefined boundary set during 2004.} For the others, it has been much earlier -- a fraction, although small, from 7-10 bin reaches this level even in 1998. This observation leads us to believe that proposed $C^3$-index has the capability of predicting future stars in advance.
 

\section{Discussion and Future Work}
In our present work, we proposed a PageRank based multi-featured author ranking metric, \emph{$C^3$-index}, that we expect would resolve some of the limitations that popular author ranking strategies such as h-index and its variants (g-index, $\hbar$-index, etc.) usually suffer from. \textcolor{black}{One of the serious problems that we addressed here is the difficulty in ranking low-profile authors who are majority in number. The difficulty arises due to the fact that h-index and its variants produce integral scores spanning over a very low bounding range.} The PageRank based strategy has been shown to overcome this problem.

The next issue we handled is the selection of features to devise the ranking strategy. We chose three features -- the quality of citations received by the papers published by the concerned authors, the quality of citations received by the author from his/her peers, and the quality of coauthors he/she had worked with. All these features were collectively represented in the form of a multi-layer bibliographic network, which was further used for ranking the authors. There are three components in the expression for computing $C^3$-index, viz, ACI-score, PCI-score and AAI-score, each connected with one of the features mentioned above. We observed that popular author ranking indices like h-index and its variants have very high correlation with ACI-score component, but have significantly low correlation with the other two components. We may infer from this information that our proposed score carries more information about an author than h-index and its variants.

The third issue that has been addressed here is to find the relation of aforementioned components of $C^3$-index with the profile of the scientific authors. Temporal plots of $C^3$-index against h-index across the years reveal that the large fraction of authors having higher AAI-score component at a particular time attain larger values of h-index in future than authors having lower AAI-score. In other words, AAI component carries some indicator of future performance of an author within it. Interpreting differently, we may claim that $C^3$-index ranks an author not only on the basis of his/her present but also on his/her future prospect. 


\textcolor{black}{The fourth issue is to extend further the scope of future author performance prediction through the proposed strategy.} We observed that $C^3$-index reveals the future outreach of a good fraction of the selected authors much earlier than the actual time they reached that milestone. \textcolor{black}{This may indicate an additional scope of application for the proposed strategy than mere ranking of authors based on their present performance.} We shall also check the results on other datasets from different domains such as Physics, Biology to strengthen our claims.



\if{0}
\bibliographystyle{spmpsci}      
\bibliography{Scientometrics_Ref}   
\fi


\end{document}